\documentclass[useAMS,usenatbib]{mn2e}
\usepackage[totalwidth=500pt,totalheight=690pt]{geometry}
\usepackage[english]{babel}
\usepackage{natbib}
\usepackage{amsmath}
\usepackage{graphicx}
\usepackage{amssymb}

\newcommand{\Ha}{${\rm H\alpha}$\ }
\newcommand{\HFratio}{${\rm H\alpha/}$FUV\ }
\newcommand{\HII}{H\thinspace{\sc ii}\ }

\newcommand{\msun}{M_{\rm \odot}}
\newcommand{\lfuv}{$L_{\rm FUV}$\ }
\newcommand{\lha}{$L_{\rm H\alpha}$\ }

\begin{document}

%-*-LaTeX-*-
% Copied from gmorris
% Note that some of these call others, eg \kmps uses \km.

% journals
\newcommand{\apj}{ApJ}
\newcommand{\mnras}{MNRAS}
\newcommand{\nat}{Nat}
\newcommand{\physrevB}{Phys. Rev. B}
\newcommand{\araa}{ARA\&A}                % "Ann. Rev. Astron. Astrophys."
\newcommand{\aap}{A\&A}                   % "Astron. Astrophys."
\newcommand{\aaps}{A\&AS}                 % "Astron. Astrophys. Suppl. Ser."
\newcommand{\aj}{AJ}                      % "Astron. J."
\newcommand{\apjs}{ApJS}                  % "Astrophys. J. Suppl. Ser."
\newcommand{\pasp}{PASP}                  % "Publ. Astron. Soc. Pac."
\newcommand{\apjl}{ApJ}                   % letter at ApJ
\newcommand{\pasj}{PASJ}
\newcommand{\gca}{Geochim. Cosmochim. Acta}
\newcommand{\physrep}{Phys. Rep.}
\newcommand{\aapr}{A\&A Rev.}
%\newcommand{\apss}{Ap\&SS} 		  % "Astrophys. Space Sci."

% Length
\newcommand{\Mpc}{\rm\thinspace Mpc}
\newcommand{\kpc}{\rm\thinspace kpc}
\newcommand{\pc}{\rm\thinspace pc}
\newcommand{\km}{\rm\thinspace km}
\newcommand{\m}{\rm\thinspace m}
\newcommand{\cm}{\rm\thinspace cm}
\newcommand{\pix}{\rm\thinspace pixel}
\newcommand{\cmps}{\hbox{$\cm\s^{-1}\,$}}
\newcommand{\cmpssq}{\hbox{$\cm\s^{-2}\,$}}
\newcommand{\cmsq}{\hbox{$\cm^2\,$}}
\newcommand{\cmcu}{\hbox{$\cm^3\,$}}
\newcommand{\pcmcuK}{\hbox{$\cm^{-3}\K\,$}}
\newcommand{\ppixsq}{\hbox{$\pix^{-2},$}}
\newcommand{\kpcsq}{\hbox{$\kpc^{2}\,$}}
% Length

% Time
\newcommand{\yr}{\rm\thinspace yr}
\newcommand{\Gyr}{\rm\thinspace Gyr}
\newcommand{\Myr}{\rm\thinspace Myr}
\newcommand{\s}{\rm\thinspace s}
\newcommand{\ks}{\rm\thinspace ks}
\newcommand{\mus}{\hbox{$\mu\s\,$}}
% Time

% Frequency
\newcommand{\GHz}{\rm\thinspace GHz}
\newcommand{\MHz}{\rm\thinspace MHz}
\newcommand{\Hz}{\rm\thinspace Hz}
% Frequency

% Temperature
\newcommand{\K}{\rm\thinspace K}
% Temperature

% Pressure
\newcommand{\Kpcmc}{\hbox{$\K\cm^{-3}\,$}}
% Pressure

% Mass
\newcommand{\g}{\rm\thinspace g}
\newcommand{\gpcm}{\hbox{$\g\cm^{-3}\,$}}
\newcommand{\gpcmps}{\hbox{$\g\cm^{-3}\s^{-1}\,$}}
\newcommand{\gps}{\hbox{$\g\s^{-1}\,$}}
\newcommand{\Msun}{\hbox{$\rm\thinspace M_{\odot}$}}
\newcommand{\Msunpc}{\hbox{$\Msun\pc^{-3}\,$}}
\newcommand{\Msunpkpc}{\hbox{$\Msun\kpc^{-1}\,$}}
\newcommand{\Msunppc}{\hbox{$\Msun\pc^{-3}\,$}}
\newcommand{\Msunppcpyr}{\hbox{$\Msun\pc^{-3}\yr^{-1}\,$}}
\newcommand{\Msunpyr}{\hbox{$\Msun\yr^{-1}\,$}}
% Mass

% Energy
\newcommand{\MeV}{\rm\thinspace MeV}
\newcommand{\keV}{\rm\thinspace keV}
\newcommand{\eV}{\rm\thinspace eV}
\newcommand{\erg}{\rm\thinspace erg}
\newcommand{\Jy}{\rm\thinspace Jy}
\newcommand{\W}{\rm\thinspace W}
\newcommand{\mJy}{\rm\thinspace mJy}
\newcommand{\ergcmcups}{\hbox{$\erg\cm^3\ps\,$}}
\newcommand{\ergpcmps}{\hbox{$\erg\cm^{-3}\s^{-1}\,$}}
\newcommand{\ergpcmsq}{\hbox{$\erg\cm^{-2}\,$}}
\newcommand{\ergpcmcu}{\hbox{$\erg\cm^{-3}\,$}}
\newcommand{\ergpcmsqps}{\hbox{$\erg\cm^{-2}\s^{-1}\,$}}
\newcommand{\ergpcmsqpspA}{\hbox{$\erg\cm^{-2}\s^{-1}$\AA$^{-1}\,$}}
\newcommand{\ergpcmsqpspsr}{\hbox{$\erg\cm^{-2}\s^{-1}\sr^{-1}\,$}}
\newcommand{\ergpcmcups}{\hbox{$\erg\cm^{-3}\s^{-1}\,$}}
\newcommand{\ergps}{\hbox{$\erg\s^{-1}\,$}}
\newcommand{\ergpspmp}{\hbox{$\erg\s^{-1}\Mpc^{-3}\,$}}
\newcommand{\ergpspkpc}{\hbox{$\erg\s^{-1}\kpc^{-1}\,$}}
\newcommand{\keVpcmsqpspsr}{\hbox{$\keV\cm^{-2}\s^{-1}\sr^{-1}\,$}}
\newcommand{\WpHzpsr}{\hbox{$\W\pHz\psr\,$}}
% Energy

% Force
\newcommand{\dyn}{\rm\thinspace dyn}
\newcommand{\dynpcmsq}{\hbox{$\dyn\cm^{-2}\,$}}
% Force

% Speed
\newcommand{\kmps}{\hbox{$\km\s^{-1}\,$}}
\newcommand{\kmpspmp}{\hbox{$\km\s^{-1}\Mpc{-1}\,$}}
\newcommand{\kmpspMpc}{\hbox{$\kmps\Mpc^{-1}\,$}}
% Speed

% Luminosity
\newcommand{\Lsun}{\hbox{$\rm\thinspace L_{\odot}$}}
\newcommand{\Lsunppc}{\hbox{$\Lsun\pc^{-3}\,$}}
% Luminosity

% Misc
\newcommand{\plawnorm}{\hbox{$\rm\thinspace photons\keV^{-1}\cm^{-2}\s^{-1}\,$}}
\newcommand{\Zsun}{\hbox{$\thinspace \mathrm{Z}_{\odot}$}}

\newcommand{\gauss}{\rm\thinspace gauss}
\newcommand{\chisq}{\hbox{$\chi^2$}}
\newcommand{\delchi}{\hbox{$\Delta\chi$}}
\newcommand{\ph}{\rm\thinspace ph}
% Misc

% Angles
\newcommand{\amin}{\rm\thinspace arcmin}
\newcommand{\aminsq}{\hbox{$\amin^{2}\,$}}
\newcommand{\asec}{\rm\thinspace arcsec}
\newcommand{\asecsq}{\hbox{$\asec^{2}\,$}}
\newcommand{\sr}{\rm\thinspace sr}
\newcommand{\degword}{\rm\thinspace deg}
\newcommand{\degsq}{\hbox{$\degword^2\,$}}
% Angles

% Emission measure
\newcommand{\emm}{\hbox{$\cm^{-5}\,$}}
\newcommand{\empasecsq}{\hbox{$\emm\asec^{-2}\,$}}
\newcommand{\keVempasecsq}{\hbox{$\keV\emm\asec^{-2}\,$}}
% Emission measure

% Per something
\newcommand{\pcm}{\hbox{$\cm^{-1}\,$}}
\newcommand{\psqcm}{\hbox{$\cm^{-2}\,$}}
\newcommand{\pcmsq}{\hbox{$\cm^{-2}\,$}}
\newcommand{\pcmcu}{\hbox{$\cm^{-3}\,$}}
\newcommand{\pmpc}{\hbox{$\Mpc^{-1}\,$}}
\newcommand{\pmpccu}{\hbox{$\Mpc^{-3}\,$}}
\newcommand{\ps}{\hbox{$\s^{-1}\,$}}
\newcommand{\pHz}{\hbox{$\Hz^{-1}\,$}}
\newcommand{\pcmK}{\hbox{$\cm^{-3}\K$}}
\newcommand{\phpcmsqps}{\hbox{$\ph\cm^{-2}\s^{-1}\,$}}
\newcommand{\psr}{\hbox{$\sr^{-1}\,$}}
% Per something

% Use times for maths fonts
% \DeclareSymbolFont{operators}   {OT1}{ptmcm}{m}{n}
% \DeclareSymbolFont{letters}     {OML}{ptmcm}{m}{it}
% \SetMathAlphabet{\mathbf}{normal}{\encodingdefault}{\rmdefault}{\bfdefault}{n}%
% \SetMathAlphabet{\mathsf}{normal}{\encodingdefault}{\sfdefault}{m}{n}%
% \SetMathAlphabet{\mathrm}{normal}{\encodingdefault}{\rmdefault}{m}{n}%
% \SetSymbolFont{operators}{bold}{OT1}{ptmcm}{b}{n}
% \SetSymbolFont{letters}{bold}{OML}{ptmcm}{b}{it}
% \SetMathAlphabet{\mathbf}{bold}{\encodingdefault}{\rmdefault}{m}{n}%
% \SetMathAlphabet{\mathsf}{bold}{\encodingdefault}{\sfdefault}{b}{n}%
% \SetMathAlphabet{\mathrm}{bold}{\encodingdefault}{\rmdefault}{b}{n}%

%\thinmuskip=2.5mu
%\medmuskip=3.5mu plus 1mu minus 1mu
%\thickmuskip=4.5mu plus 1.5mu minus 1mu
%\DeclareSymbolFont{symbols}     {OMS}{cmsy}{m}{n}
%\DeclareSymbolFont{largesymbols}{OMX}{cmex}{m}{n}
%\SetSymbolFont{symbols}{bold}{OMS}{cmsy}{b}{n}
%\SetSymbolFont{largesymbols}{bold}{OMX}{cmex}{m}{n}

\title[\Ha to FUV ratios in star forming regions of nearby spirals]{\Ha to FUV ratios in resolved star forming region populations of nearby spiral galaxies}
\author[M.T. Hermanowicz, R.C. Kennicutt and J.J. Eldridge]{Maciej T. Hermanowicz$^{1}$\thanks{E-mail:
mherman@ast.cam.ac.uk (MTH)}, Robert C.
Kennicutt$^{1}$ and John J. Eldridge$^{1,2}$\\
$^{1}$Institute of Astronomy, University of Cambridge, Cambridge, CB3 0HA,  UK\\
$^{2}$The Department of Physics, The University of Auckland, Private Bag 92019, Auckland, New Zealand}

\date{Accepted 2012 month xx. Received 2012 Month xx; in original form 2012 month xx}

\pagerange{\pageref{firstpage}--\pageref{lastpage}} \pubyear{2013}

\maketitle

\label{firstpage}

\begin{abstract}
We present a new study of \HFratio flux ratios of star forming regions within a sample of nearby spiral galaxies.
We search for evidence of the existence of a cluster mass dependent truncation in the underlying stellar initial mass function (IMF).
We use an automated approach to identification of extended objects based on the SExtractor algorithm to catalogue resolved \HII regions within a set of nearby spiral galaxies.   
Corrections due to dust attenuation effects are applied to avoid artificially boosted \HFratio values. 
We use the BPASS stellar population synthesis code of \citet{BPASS2009} to create a benchmark population of star forming regions to act as a reference for our observations. 
Based on those models, we identify a zone of parameter space populated by regions that \rm{cannot be obtained with a cluster mass dependent truncation in the stellar IMF imposed}.
We find that the investigated galaxies display \rm{small} subpopulations of star forming regions falling within our zone of interest, \rm{which appears to be} inconsistent with imposing an IMF truncation at maximum stellar mass dependent on the total cluster mass.
\rm{The presence of those regions is expected in models using both stochastic and sorted sampling of the full extent of the stellar IMF.} 
This result persists after taking dust attenuation effects into account. 
We highlight the significance of stochastic effects in environments with low star formation activity and in studies describing systems associated with small physical scales.
\end{abstract}


\section{Introduction}

The stellar Initial Mass Function (IMF) is the key property characterising a  newly formed stellar population. 
It describes the underlying probability distribution of the formation of a star within a particular mass range.
Currently, the most commonly used parametrisation is a two-part power law $\xi(m) \propto m^{-\alpha_{i}}$ presented by \citet{Kroupa_IMF}, with $\alpha_1=1.3$ for $0.1\leq m/M_\odot \le 0.5$ and $\alpha_2=2.35$, identical with the slope from \citet{Salpeter}, for $0.5\leq m/M_\odot \le m^{*}_{ \rm max}$, where $m^{*}_{ \rm max}$ is the fundamental upper stellar mass limit. 
At Solar metallicity $m^{*}_{\rm max} = 150 \rm M_\odot$  is widely adopted \citep[and the references within]{Zinnecker2007}, although $120 \rm M_\odot$ is also often used.

Over the recent years a considerable amount of discussion was devoted to the question of whether the stellar IMF is indeed universal as postulated by many works \citep[e.g.][]{Kroupa_IMF}.  \citet*{Bastian_IMF_review} and the references therein provide an extensive review of the observational evidence on IMF variation. 
The Integrated Galactic IMF (IGIMF) theory \citep{Kroupa&Weidner2003,Weidner&Kroupa2005,Weidner&Kroupa2006} has been developed to explain the claims \citep[e.g][]{Prisinzano,Kalirai,Moraux} for a presence of a top-light IMF (i.e. declining more steeply within the high mass regime). 
The core of the IGIMF paradigm is that the effective galactic IMF is composed of the contributions from individual star forming events. Each of those events corresponds to formation of a cluster and a separate sampling of the adequate underlying stellar IMF.
Note that this view encompasses both clusters that remain bound and that disperse to form stellar associations.

\rm{The semi-analytic implementation of the IGIMF theory is founded upon two underlying assumptions.
First, the maximum mass of a cluster formed in a galaxy ($M_{\rm ecl,max}$) is taken to be a function of the star formation rate (SFR) of the galaxy \citep*{WeidnerKroupaLarsen2004}.  
Second, the maximum mass of the star $m_{ \rm max}$ that can be formed within a newly formed cluster of mass $M_{\rm ecl}$ is dependent on $M_{\rm ecl}$ itself.  
This semi-analytical relation was presented in \citet{Weidner&Kroupa2006} and was described in further detail by \citet*[afterwards PA07]{Pflamm07}.  
We will be referring to it as the PA07 relation.}

Two compilations of published Milky Way clusters datasets attempted to demonstrate this relationship. 
\citet{Maschberger2008} analysed a sample of Milky Way clusters and found no significant departures from the stochastic sampling of the full extent of the IMF in low mass clusters, as described earlier by \citet{Elmegreen2006}. No alternative sampling methods have been investigated, though.
\citet*{Weidner2010} presented a similar compilation of a larger cluster sample extending to higher masses, containing observational evidence supporting the PA07 relation. 
Those authors concluded that a relationship between $M_{\rm ecl}$ and $m_{\rm max}$ is a physical constraint rather than a statistical effect. They also point out that investigating a wider cluster mass range is necessary for distinguishing between different sampling methods.
\citet{Elmegreen2006} interprets the relationship between $M_{\rm ecl}$ and $m_{\rm max}$ as a statistical connection originating from the size of sample effects. 

Rather than attempting to verify the PA07 relation directly, one can search for its manifestations on galactic and sub-galactic scales.  
The semi-analytic IGIMF paradigm predicts a divergence of \Ha and far ultraviolet (FUV) fluxes of galaxies with low SFRs \citep*{Pflamm09}. 
This is because below a certain SFR threshold formation of stars massive enough to have significant ionizing flux becomes impossible. 
\citet{Pflamm09} predict that this should translate into a discrepancy between the star formation rates recovered from \Ha and FUV calibrations. 
\rm{An attempt by \citet{Lee2009} to observe this trend for integrated galactic fluxes has confirmed a systematic turnover in \HFratio ratio at low SFRs, but also showed that the origin of the trend was inconclusive.}

\rm{The $M_{\rm ecl}$ - $m_{\rm max}$ relationship presented by PA07 has been widely overinterpreted as an imposition of a strict cluster mass dependent truncation in the stellar IMF. 
\citet{Pflamm09} predictions have sparked further investigations into the potential presence of a sharp IMF cutoff $m_{\rm max} < m^{*}_{\rm max}$ at low cluster masses. 
\citet*{Fumagalli2011} and \citet{Weisz2012} demonstrated discrepancies between the observations and predictions from a variable IMF truncation combined with the $M_{\rm ecl,max}$ - SFR relationship \citep{WeidnerKroupaLarsen2004}. 
\citet{Eldridge2012} has found similar discrepancies when using a truncated IMF in absence of a $M_{\rm ecl,max}$-SFR dependence.}

\rm{Previous studies of integrated galactic properties question the validity of using a truncated IMF models inspired by the semi-analytic IGIMF.
However, they cannot determine whether the inconsistencies arise due to the implementation of the IMF truncation, the $M_{\rm ecl,max}$ - SFR relationship or both.  
The aim of this study is to verify the first of those options by investigating star formation on subgalactic scales.}
To accomplish this, we perform a search within a sample of nearby spiral galaxies for a population of star forming regions that are inconsistent with \rm{the presence of a cluster-mass-dependent upper truncation to the IMF.}
We use the \textsc{Binary Population And Spectral Synthesis} (\textsc{BPASS}) stellar population synthesis code \citep{BPASS2009} to extend the \HFratio flux divergence predictions of \citet{Pflamm09} to \rm{the regime of individual clusters with various approaches to IMF sampling}.
This study is complementary to work of \citet{Corbelli2009} on M33, \citet{Calzetti2010} on NGC 5194 and \citet{Koda2012} on the extended UV (XUV) disc of NGC 5236.

\section{Data sample}\label{section_sample}

\begin{table*}
\begin{minipage}{200mm}
\begin{tabular}{|l|l|l|c|c|c|c|c|c|}
\hline
Name & RA & Dec	& distance & type & $E(B-V)$ &	$\log(L_{\rm H\alpha})$	&	$\log(L_{\rm FUV})$	&	telescope\\
 &	(J2000) &	(J2000) & [Mpc]	& &  [mag]	& $\rm[erg\ s^{-1}]$ &  $\rm[erg\ s^{-1}]$	& 		\\
\hline
NGC 300	        & 00:54:53	& -37:41:04	& 2.0$^{(1)}$	& SA(s)d	& 0.0130	& 40.18		& 42.46		& Danish 1.54 m$^{(5)}$\\
NGC 628  	& 01:36:42	& +15:47:01     & 8.6$^{(2)}$   & SA(s)c	& 0.07	        & 40.87		& 43.02		& CTIO 1.5m$^{(6)}$\\	
NGC 2403	& 07:36:51	& +65:36:09	& 3.22$^{(1)}$	& SAB(s)cd	& 0.04	        & 40.78		& 42.83		& KPNO 2.1m$^{(6)}$\\											
NGC 3031	& 09:55:33      & +69:03:55     & 3.63$^{(1)}$  & SA(s)ab	& 0.08	        & 40.77		& 42.76		& KPNO Schmidt$^{(7)}$\\
%NGC 5194	& 13:29:53	& +47:11:43	& 7.6$^{(2a)}$	& SA(s)bc pec	& 0.04	        & 41.28		& 43.34		& KPNO 2.1m$^{(6)}$\\
NGC 5236	& 13:37:01	& -29:51:56	& 4.61$^{(3)}$	& SAB(s)c	& 0.066	        & 41.25		& 43.21		& CTIO 1.5m$^{(8)}$\\
NGC 5457	& 14:03:13	& +54:20:56	& 6.7$^{(1)}$	& SAB(rs)cd	& 0.01	        & 41.33	        & 43.61		& KPNO Schmidt$^{(7)}$\\
NGC 7793	& 23:57:50	& -32:35:27	& 3.91$^{(4)}$	& SA(s)d	& 0.02	        & 40.58		& 42.67		& CTIO 1.5m$^{(6)}$\\							
\hline

\end{tabular}
\end{minipage}

\caption{Overview of the galaxies sample. \Ha luminosities taken from \citet{Kennicutt2008}. $L_{\rm FUV}=\lambda_{\rm FUV}F_{\rm FUV}$ values taken from \citet{Lee2011}. References indicated in the ``telescope" column are the sources for \Ha data used in the analysis. \textit{References:} $^{(1)}$\citet{Freedman2001}; $^{(2)}$\citet{NGC628_distance}; $^{(3)}$\citet{Saha2006};  $^{(4)}$\citet{NGC7793_distance};  $^{(5)}$\citet{Larsen1999};   $^{(6)}$\citet{SINGS};  $^{(7)}$\citet*{Hoopes2001}; $^{(8)}$\citet{SINGG};}
\label{data_sample}

\end{table*}
% $^{(2a)}$\citet{M51_distance} skipped

For the purposes of this experiment we have selected a sample of seven nearby nearly face-on spiral galaxies for which  ground-based $\rm H\alpha$, FUV, NUV and \textit{Spitzer Space Telescope} MIPS $\rm{24\mu m}$ data are available.
The galaxies were selected to span a narrow distance range.  
This is because more distant galaxies are not resolved enough to infer the properties of small enough star forming regions.
On the other hand, the assumption of spatial coincidence of the sources of ionizing radiation and \Ha emission breaks down in better resolved nearby galaxies 
 
 Table \ref{data_sample} provides an overview of the properties of these galaxies.  The galaxies' positions, morphological types and $E(B-V)$ values were obtained directly from the NASA/IPAC Extragalactic Database (NED). 
In most cases we adopt the same distances as \citet{Lee2011} did in their study.  The one exception is NGC 628, where we use more recent measurements based on the observed luminosity functions of planetary nebulae.
Throughout the analysis we adopt the distance uncertainties as quoted in the relevant papers. The references are listed in the caption of Table \ref{data_sample}. 

The MIPS $\rm{24\mu m}$ data are  taken from two multi-wavelength datasets: Local Volume Legacy survey \citep[LVL,][]{LVL} and Spitzer Infrared Nearby Galaxies Survey  \citep[SINGS,][]{SINGS}. 
Most of the narrow-band \Ha images are obtained from sets of ancillary observations to the two abovementioned surveys.
In the case of NGC 300 we used data from the Survey for Ionization in Neutral Gas Galaxies \citep[SINGG,][]{SINGG}.  
\rm{For each of the galaxies} we selected the highest quality \Ha data available. 
The references for \Ha images are quoted in the description of Table \ref{data_sample}. 
Note that due to the large mismatch \rm{between the} point spread functions (PSFs) of FUV and infrared (IR) data, high angular resolution of the \Ha data is not necessarily a significant advantage.

We used \textsc{GAIA} software \citep{GAIA} to manually remove star subtraction artefacts from the narrow-band \Ha images using the inbuilt patching functions.
We typically used circular apertures with a third order fit (noise estimate scaling factor of $1.0$) to a background annulus in the patching procedure.
The annuli's radii were selected in order to avoid neighbouring objects.
\textsc{GAIA} fits a single background surface across all the regions selected.  
Hence, in order to optimize the quality of the process, patching was performed for each region individually.
In more complex cases, such as crowded regions or presence of artefacts associated with prominent diffraction spikes in the continuum image, appropriate patching region shapes were selected to compensate.  
In such situations, we were often forced to abandon the use of background annuli in favour of individually assigned sets of background regions.
We find that as a result of patching, the quality of object identification is significantly improved.
We exclude the regions near the patched out objects from the output of our search algorithm.

\begin{figure}
\centering
\includegraphics[width = 1.0\columnwidth]{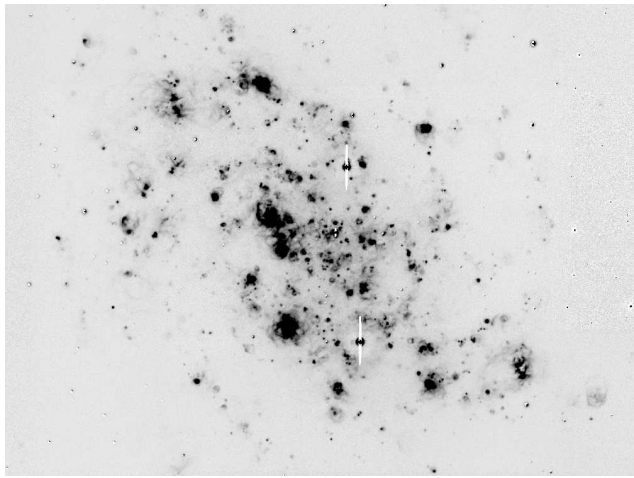}
\caption[NGC 2403: \Ha image]{NGC 2403: continuum subtracted \Ha emission prior to patching out of artefacts.}
\label{NGC2403}
\end{figure}

Where available, we retained the published continuum subtractions. This was not the case with NGC 2403 ancillary data from SINGS, where astrometry of the published R-band continuum image was incorrect. 
We used  \textsc{GAIA} software to rederive astrometry for both images by matching the stars to the Guide Star Catalog II \citep{GSC_2}. 
We then subtracted the R band continuum image, scaled by a uniform factor $k_R$, off the narrow band \Ha image to correct for the presence of continuum. 
R-band image has been converted to the pixel scale of \Ha data and convolved with a 2D Gaussian kernel with FWHM $=3.0$ pix (to match \Ha image PSF) prior to performing the subtraction.
The value of $k_R$ has been derived by performing photometry of a sample of 20 manually selected stars on both images and adopting the median scaling.
However, we note that there is a second order effect due to mosaicking that can contribute errors of the order of $5$ per cent in flux.
The final \Ha map  is displayed in Figure \ref{NGC2403}. 
In the case of NGC300 we used the \textit{astrometry.net} web-based program \citep{astrometry.net} to provide an astrometric calibrations for the images of \citet{Larsen1999}.

The FUV images are taken from the LVL survey \citet{LVL,Lee2011}.
Most of these observations were originally presented in the  \textit{Galaxy Explorer Satellite} (\textit{GALEX}) atlas of nearby galaxies by \citet{GALEX_atlas}, which provides the details of the image processing pipelines used. 
We cross-checked our flux calibrations against the integrated galaxy values of \citet{Lee2011} and found excellent agreement.
Foreground extinction correction was applied adopting the $E(B-V)$ values from \citet*{Schlegel} and the Milky Way extinction curve of \citet{Pei1992}.

\section{Methodology}

This section describes the methodology adopted in our study.  First, a summary of our approach to object identification is given. Second, the method of correcting for  dust attenuation is described.

\subsection{Object identification and photometry} \label{section:detection_method}

In order to identify our regions of interest, we run \textsc{SExtractor} \citep{SExtractor} package using \Ha images of galaxies as reference to select the regions of recent star formation. 
The code first fits a global background profile using  $\kappa.\sigma$-clipping, similar to that employed in \textsc{DAOPHOT} \citep{DAOPHOT}.
 \textsc{SExtractor} initially identifies objects \rm{to be} groups of pixels with values exceeding \rm{a set} threshold above this background fit.

Multiple isophotal deblending is applied to the initial detections to evaluate significance of the individual peaks.
The expected contributions from bivariate Gaussian fits to deblending objects are then used to allocate contended pixels.
\textsc{SExtractor} then eliminates spurious detections within wings of shallow profile objects by evaluating the contribution to the mean surface brightness from the gaussian extrapolation of neighbouring objects. If the corrected surface brightness exceeds the detection threshold, the object is accepted into the catalogue.
Typical SExtractor configuration parameters used in our study are listed in Table \ref{SEx_table}. They were chosen to maximize the detection efficiency of the population of faint objects.

The program is run in single image mode on \Ha data and in dual image mode on the remaining three bands.  Objects identified in this way are referred to throughout the paper as ``\Ha selected''.  
A second set of runs is performed using FUV image as the reference, producing what we refer to as the ``FUV selected'' sample.
This paper focuses on the results from the \Ha selected sample.

\begin{table} 
\begin{tabular}{l | l | l}
\hline 
parameter name & set 1  & set 2  \\
\hline
DETECT\_MINAREA  &  10 & 35\\   
THRESH\_TYPE     &  RELATIVE &  RELATIVE\\      
DETECT\_THRESH   &  5.0 &  5.0\\
ANALYSIS\_THRESH &  5.0  &  5.0  \\         
DEBLEND\_NTHRESH &  64  &  64     \\              
DEBLEND\_MINCONT &  0.00001 &  0.00001  \\       
CLEAN           &  Y    &  Y       \\            
CLEAN\_PARAM    &   10.0  &   10.0     \\    
BACK\_SIZE      &   35  &   100 \\        
BACK\_FILTERSIZE &  6  &  6    \\   
BACK\_TYPE       &  AUTO  &  AUTO  \\     
BACKPHOTO\_TYPE  &  LOCAL &  LOCAL  \\   
BACKPHOTO\_THICK &  10 &  24\\
\hline

\end{tabular}
\caption[\textsc{SExtractor} input.]{Adopted values of \textsc{SExtractor} input parameters.  Set 1 has been used for analysis in native FUV resolution and for the analysis in \Ha resolution in cases of images taken with the CTIO 1.5m and KPNO Schmidt telescopes, due to their  comparatively poor resolution.  Set 2 was adopted for higher resolution \Ha images.}
\label{SEx_table}

\end{table}

Aperture photometry is performed on the list of detections by applying the \textsc{APER} procedure of the \textsc{IDLASTRO} package.  
We use the \textsc{SExtractor} defined isophotal apertures with signal-to-noise threshold of 5.0 and an appropriate minimum area. 
We mask out regions that are identified outside the galaxies' disks, within their bulges and near the artefacts of continuum subtraction that were removed from \Ha images during the patching stage. 
A set of circular apertures is used to allow for spatial offsets between the FUV and \Ha emission.  
The aperture selected for any given region is the smallest of the set of available apertures that has an area larger Wthan the isophotal area fitted to the region.

We use two sets of ten apertures each. For each galaxy, we choose one of the two aperture sets, the choice being dependent on the angular sizes of discernible \HII regions due to the host galaxy's distance and telescope resolution. 
Smallest aperture diameter in ``set 1'' is 4 pixels and the sizes increase in 2 pixel steps until the maximum of 24 pixels.
``Set 2'' apertures range from 10 to 55 pixels with a 5 pixel increment in diameter.  
``Set 1'' has been used for NGC 3031 and NGC 5457, which have low resolution \Ha data from the KPNO Schmidt telescope.  Remaining galaxies had ``set 2'' applied in the analysis pipeline.  
  
All the images have been rebinned and realigned to the pixel scale of the relevant \Ha image using a modified \texttt{frebin} IDL routine, with an added flux conservation option. 
Aperture corrections have been derived by using the same photometric approach on rebinned FITS images of the PSFs of relevant instruments.  
The PSF information was obtained directly from the websites of \textit{GALEX} and \textit{Spitzer} missions.
\textsc{ATV} package  \citep{ATV} was used for quick check photometry and result verification, since it allows for direct application of the \textsc{APER} routine.

\subsection{Dust attenuation correction}

Dust attenuation produces a strong effect on the locus of points on \HFratio diagrams.
The data points are displaced along a vector with a direction fixed by the adopted attenuation curve and length determined by the magnitude of the attenuation. 
It is universally acclaimed that FUV is more significantly attenuated than the \Ha nebular line \citep[e.g.][]{Calzetti2000}.  Therefore, any study which fails to correct for this effect will find that their objects  have artificially boosted values of the \HFratio ratio.  
\citet{Koda2012}  accounted for this effect using a uniform attenuation of $A_V = 0.1$ mag.  
This approach was possible because the XUV disc environment they investigated has low dust content. 
Therefore, spatial variations in attenuation would not affect the result of their work significantly.

In this study, we are dealing with regions within the optically bright discs of spiral galaxies.
Adopting the approach of \citet{Koda2012} would result in shifting the locus of all the data points in our figures along the overplotted arrows by a fixed amount.  
However, a significant spatial variation in the attenuation values within a disk of a spiral galaxy is expected \citep[e.g.][]{Roussel2005}.
The distribution of dust tends to be concentrated around the sites of star formation \citep{Prescott2007}.
Therefore, a uniform attenuation value derived from net galactic fluxes will most likely underestimate the real attenuation in our objects.

In view of this issue we turn to estimating individual attenuation values for each of the identified regions.  
We adopt a fixed dust attenuation curve for all the regions, taken from \citet{Calzetti2001}, yielding the relation $A_{\rm FUV} = 1.82\ A_{\rm H\alpha}$.  

This scaling relation incorporates both the attenuation curve values and the difference between the attenuation of stellar light and nebular emission predicted from the typical star forming region geometry.
The dust corrections are calculated using the relation between FUV and MIPS 24 $\mu m$ fluxes using the calibration from Section 4.3 of \citet{Hao2011}, yielding the relation in Equation \eqref{afuv}. \rm{Throughout the paper we use $L_{\rm FUV}=\lambda_{\rm FUV}F_{\rm FUV}$.}

\begin{equation}
A_{\rm{FUV}} = 2.5\log_{10}(1+(3.89 \frac{L_{\rm{24 \mu m}}}{L_{\rm FUV}}))  \label{afuv}
\end{equation}

We are implicitly assuming that both the nebular emission and the UV-bright stellar populations within our apertures  are attenuated by the same dust distribution. 
The adoption of the \citet{Calzetti2001} attenuation relation imposes the geometry of clumpy dust shell surrounding a stellar population mixed with ionized gas. 
A caveat to keep in mind is that the \rm{relative} geometry of dust, gas and stars in the small star-forming regions targeted in our study can assume arbitrary configurations leading to a degree of stochasticity in the relative attenuations of starlight and nebular \Ha emission.
However, \citet{Calzetti2005} suggests that this model serves as a good approximation for small HII regions.  
The question of selecting appropriate dust attenuation law is explored in more depth in Section \ref{dust}.

\section{Theoretical models}

In order to predict the observational signatures of the \rm{IMF truncation}, one has to generate a set of benchmark model populations of clusters.  
A code capable of populating the IMF stochastically and considering discrete stars has to be used in order to provide reliable predictions within the regime of low-mass clusters.  
\rm{We note that although we use the term ``clusters'', we do not model the dynamics of those objects. In physical sense,they may either be bound clusters or unbound stellar associations.} 

\subsection{Model description}\label{section:models}

In this study we use the \textsc{Binary Population And Spectral Synthesis} (\textsc{BPASS}) code by \citet{BPASS2009} to provide the benchmark \rm{models}.
\textsc{BPASS} is a stellar population synthesis code based on a library of 15,000 stellar evolution models computed using the Cambridge  \textsc{STARS} code.  
The models used are described in detail by \citet*{STARS_binary}. 
\citet{Eldridge2012} recently used the package to investigate the effects of stochasticity and interpret the attempts of observational verification of IGIMF presented recently by \citet{Lee2009,Lamb2010,Calzetti2010}. 
We refer the reader to $\S2$ of \citet{Eldridge2012} for a more detailed description of the methods used to create the models\rm{, including the implementation of binary evolution}.
The relevant aspects of the model generation will be covered briefly here.

\begin{figure*}
\begin{minipage}{180mm}
\centering
\includegraphics[width = 1.0\columnwidth]{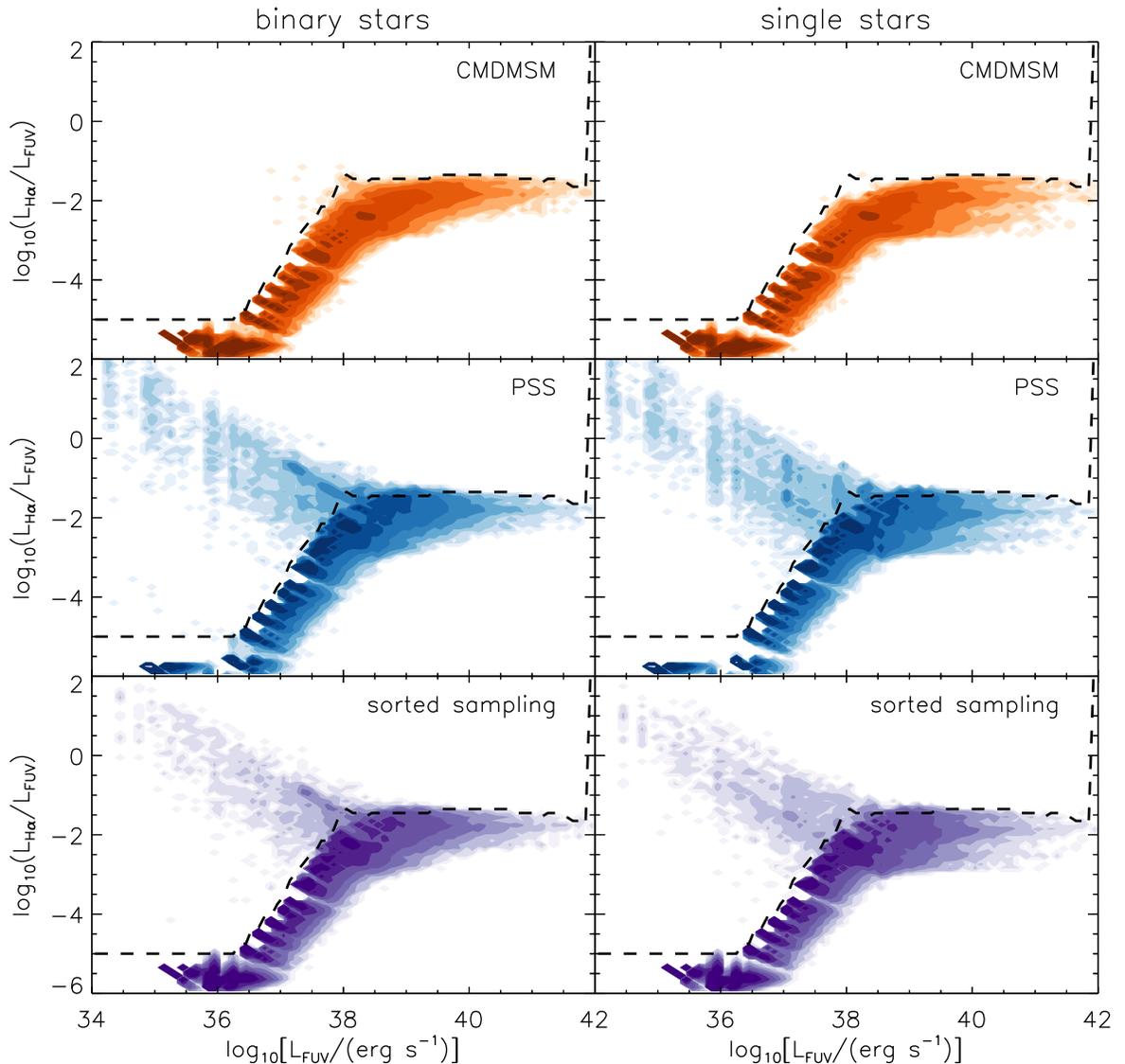}
\end{minipage}
\caption[Model populations]{\rm{Number density contour plots of BPASS model populations used in this study presented in the \HFratio versus \lfuv plane.  \textit{Orange}: cluster mass dependent maximum stellar mass (CMDMSM) models corresponding to the truncated IMF; \textit{blue}: pure stochastic sampling (PSS) models; \textit{purple}: models generated with sorted sampling, preferred by the IGIMF paradigm (see Section \ref{section:models} for details). Contours are plotted every 0.5 dex in number density. Binning of 0.1 dex adopted in XY. Left-hand side column illustrates models with binary evolution effects included. Right-hand side column illustrates models with single stars only. Note that the majority of models at low FUV luminosities($<10^{39}$\ergps) follow the turnoff consistent with the IMF truncation. Points above the dashed line are considered to be violating the truncated IMF condition.}}
\label{fig:models}
\end{figure*}

We opted for the use of this package for the following reasons. 
First, it takes binary evolution effects into account while computing the model stellar populations, setting it apart from the SLUG code used by \citet{Fumagalli2011}.
In light of the recent observational evidence \citep[e.g.][]{Pinsonneault,Kobulnicky,Kiminki} it becomes apparent that binary fraction among massive young stars is very high.
The work of \citet{Sana2012} finds an interacting binary fraction for O stars of $69$ per cent. 
Therefore, the effect of interacting binaries cannot be neglected if we are after an accurate representation of the properties of young stellar clusters.

Second, the code produces the spectral energy distributions (SEDs) of the populations by integrating over SEDs of discrete stars comprising a particular cluster.  
This approach makes the code perfectly suited for an accurate representation of the stochastic effects, which become very significant at low cluster masses.
\textsc{BPASS} has several advantages over the models of \citet{BC2003} which were used by \citet{Calzetti2010}. Already mentioned are the stochastic IMF sampling and the inclusion of interacting binary stars. 
One more important difference is the treatment of Wolf-Rayet (WR) stars. 
In \citet{BC2003} WR stars are assumed to have atmospheres similar to white dwarfs. 
This incorrectly estimates their contribution to the ionizing flux. \textsc{BPASS} uses the detailed grid of calibrated WR atmospheres, the Potsdam Wolf-Rayet models presented by \citet*{Potsdam1} and \citet{Potsdam2,Potsdam3}. 
This allows us to have greater confidence in the accuracy of estimating the ionizing contribution of WR stars in our stellar population models.

In this analysis we use models at Solar metallicity with the metallicity mass fraction of $Z_\odot=0.02$.  The clusters included in our models span an age range of 1-8 Myr.
This choice is motivated by the fact that we are expecting the objects that would violate the $m_{\rm max}$ limit to be among the youngest clusters within their host galaxies, with ongoing star fromation.  After consulting published metallicities of \citet{Moustakas2010} we expect $Z_\odot$ to be a good representation for the galaxies in our sample. 

The models cover a mass range from $0.1$ to $120 M_{\rm \odot}$ for single stars.  
This upper cutoff in the model grid is imposed by the high mass-loss rates during the main-sequence phase at Solar metallicities \citep{Vink2011}. 
\rm{For binary populations, the available initial masses are $5$ to $120 M_{\rm \odot}$. We include every integer mass up to $20 M_{\rm \odot}$, at which point the model spacing increases to $5 M_{\rm \odot}$ up to $60 M_{\rm \odot}$. Higher mass models are computed at 70, 80, 100 and $120 M_{\rm \odot}$.
We note that some binary mergers within the models can produce stars with masses in excess of $200 M_{\rm \odot}$.
Single star models have a similar grid of models, with every integer mass up to $40 M_{\rm \odot}$. To make an unbiased comparison between the single and binary populations we assume the same IMF for the primary star masses as the single stars. We then select the secondary stars mass from a flat mass ratio distribution where, $q=\frac{M_2}{M_1}$. Such a mass ratio distribution does not affect the overall IMF of  the population. The model grid in mass ratio have values of $q=0.1$, 0.3, 0.5, 0.7 and 0.9. In addition we select the initial separation of the binary from a flat distribution in $\log$ separation. The model grid ranges from an initial separation of $10$ to $10^4 R_{\odot}$ in steps of 0.25 dex. Both distributions are similar to those inferred observationally by \citet{Sana2012}. Our assumed initial distribution leads to approximately two-thirds of our binary systems experiencing interactions. For the evolution of the secondary in each model we select the most likely result after the kick of the first supernova as the evolutionary path it will follow.
This is estimated using the method of determining the binary systems' survival of the first supernova event described in \citet{EldridgeLangerTout}.}

\Ha fluxes of clusters were derived by calculating the number of ionizing photons emitted at wavelengths shortwards of $912$\AA.
We assume standard Case B recombination \citep{Osterbrock_book} so that each $10^{11.87}$ ionizing photons give rise to $1$\ergps of \Ha flux.
No flux leakage is assumed i.e. all the ionizing photons are absorbed by gas, with no corrections for the presence of dust.
FUV flux densities at $\lambda_{\rm FUV}$ were calculated directly from stacking stellar spectra in each cluster.

In order to provide a benchmark for our observations, we generated populations of model clusters using three methods.
The number density  contour plots of the resulting models are illustrated in Figure \ref{fig:models}. 
The left column presents models with binary fraction of 100$\%$, the right column presents models with no binary systems.  
In all cases we adopt the broken power-law IMF of \citet{Kroupa_IMF} with an absolute maximum initial stellar mass $m^{*}_{\rm max}=120 M_{\rm \odot}$.
The cluster populations are generated in the following manner, as described in \citet{Eldridge2012}.
First, a random cluster mass between $10$ and $10^6 \msun$ is drawn from a power law cluster IMF with a slope of $-2$ \citep{deGrijs2003,Lada&Lada}.
The cluster is then randomly filled with stars from the Kroupa IMF, truncated at mass-dependent $m_{\rm max}$ or $m^{*}_{\rm max}$, whichever is lower. 
The stars are added to the cluster until the total mass of the stars exceeds the target $M_{\rm ecl}$. 
The last star added is then retained only if the mass of the cluster obtained in this way is closer to the value of $M_{\rm ecl}$ drawn from the cluster IMF.
If the total stellar mass was closer to target $M_{\rm ecl}$ without including the final star, it is removed from the cluster. 
This corresponds to the constrained sampling described in \citet{Weidner&Kroupa2006}.

In the first set of models, we impose an upper mass truncation $m_{\rm max}$ in the cluster-mass dependent form \rm{taken from} PA07.
This mirrors previous applications of the truncated IMF by \citet{Fumagalli2011,Koda2012} and others.
We refer to this as the cluster mass dependent maximum stellar mass (CMDMSM) method. Constrained sampling is used. The resulting models are illustrated by the top pair of plots in Figure \ref{fig:models}.

\rm{The second set of models is referred to as pure stochastic sampling (PSS). We adopt a mass limit $m_{\rm max}=M_{\rm ecl}$, equivalent to allowing star of any mass within our model range to form within a cluster, provided it does not exceed the mass of the cluster itself. Constrained sampling is used. The resulting models are illustrated by the middle pair of plots in Figure \ref{fig:models}.}

\rm{The final set of models is drawn using the same $m_{\rm max}=M_{\rm ecl}$ constraint as the PSS population. 
However, instead of constrained sampling, we apply the sorted sampling preferred by \citet{Weidner&Kroupa2006}.
The difference is that upon reaching the target $M_{\rm ecl}$, the decision between rejection or retention is made for the most massive star in the cluster rather than the last one to be drawn. The resulting models are illustrated by the bottom pair of plots in Figure \ref{fig:models}.}

The synthetic model contours do not cover the parameter space in a continuous manner.  
This is because we use a discreet grid of initial masses and binary parameters. 
We refrain from interpolating between the grid since non-linearity of stellar evolution becomes an even greater issue than normally in a binary setup.
\rm{Instead, for each star drawn from the IMF we use the closest mass stellar model.
Throughout the rest of the paper we use smoothed population contours to improve clarity and alleviate the effects of a discrete stellar model grid. }

\subsection{Discriminating between the methods of populating the IMF}

The most striking feature in Figure \ref{fig:models}. is that CMDMSM models display a prominent downturn in the \HFratio ratio at low \lfuv \ values, mirroring the galaxy-wide predictions of \citet{Pflamm09}. 
As explained before, this effect is a result of the lack of massive stars with significant output of ionizing flux.

The same feature is manifested within the two remaining populations as well. 
This is because despite the lack of an IMF truncation, drawing a massive star from the IMF when forming a low-mass cluster is still unlikely. 
Therefore, the majority of model low-mass clusters will have no massive stars. 
Hence, their \HFratio ratio will drop in a similar manner as in the CMDMSM case.

\rm{Both PSS and sorted sampling realizations display a high \HFratio ratio branch. 
Those are the low-mass clusters ($M_{\rm ecl} < 10^{3.5} M_{\odot}$) which contain massive ($> 20 M_{\odot}$) stars with high output of ionizing photons. The locus of this feature is similar in both sets of models. This branch is populated by a much smaller number of objects than the turnover branch.  Sorted sampling predicts a lower relative population within this area.  }

Note that such a bimodality is expected \citep{Cervino2003} in cases where the number of stars contributing to an observable is small ($< 10$).  
The bimodality persists in BPASS models both at different metallicities and with binary evolution switched off.  
Removing binaries results in an even sharper distinction between the two branches.

The dashed line in Figure \ref{fig:models} represents a candidate selection cut. 
\rm{Any observed regions placing above this line are expected to be incompatible with having a truncated stellar IMF.}
 Note that there is a handful of CMDMSM models that lie above this selection curve as well.  
These are the models harbouring massive stars formed via binary mass transfer. 
They are outnumbered by at least two orders of magnitude by the PSS realizations in this regime though.  
Those objects are not present in the single star model realizations.

In Figure \ref{fig:models} we see that the differences between single star and binary populations are minor. 
The effect of binary evolution is to rejuventate a population, producing more massive stars at later times. 
This can be seen by examining the spread of the cluster distribution at the high FUV luminosity end. 
The single-star population display a greater spread. 
This is due to the ageing of cluster models causing a drop in \Ha ionising flux originating from a dwindling population of massive stars. 
The binary systems are able to provide the same \Ha luminosity for a longer time, which is manifested as the a slight boost of the observed \HFratio flux ratio. 
The effect of binaries can also be seen in the high \HFratio ratio arm, where the preferred locus for Wolf-Rayet stars is somewhat changed.
Furthermore, the arm is also more densely populated in the binary evolution scenarios.
This is especially visible in the CMDMSM case highlighted in the previous paragraph.

To recapitulate, we decide to adopt the high \HFratio ratio branch of the PSS distribution as the diagnostic signature of formation of stars under a stochastic scenario allowing for sampling of the stellar IMF beyond the cluster mass dependent truncation. \rm{We note that no such binary test is possible between the PSS and sorted sampling models, since they both predict the presence of the high \HFratio ratio branch.}

\section{Results}

In this section, we present the resulting descriptions of the star forming region populations within the sample of the eight investigated galaxies. 
We compare the resulting populations with the \textsc{BPASS} model predictions.
We discuss the observed influence of dust attenuations on the behaviour of our data.

\rm{Table \ref{tab:results} contains an excrept from the region catalogue for NGC 300. The full catalogues are available in a machine readable format in the online version of the paper. Regions falling outside galaxy disks and near image areas affected by artefact patching (Section \ref{section:detection_method}) are not listed. }

\begin{table*} \label{tab:results}
\begin{minipage}{200mm}
\begin{tabular}{|l|l|l|c|c|c|c|c|c|cccccccc}
\hline
ID & RA & Dec	& $d_{\rm ap}$ & $d_{\rm ap}$ & \ldots & $A_{FUV}$ &$\log(L_{\rm H\alpha, corr})$&$\log(L_{\rm FUV, corr})$ & $Y_{\rm corr}$ &note\\
   &	(J2000)&	(J2000) &[arcsec] & [pc]& \ldots & [mag] 	& $\rm[erg\ s^{-1}]$ &  $\rm[erg\ s^{-1}]$	& & &		\\
\hline

   2 &  00:54:44.00 &  -37:35:18.69  &    5.89  &  57.10 &  \ldots &   0.41 &   37.36 &   38.71  &  -1.35  &  H \\
   3 &  00:54:41.38 &  -37:35:39.49  &    5.89  &  57.10 &  \ldots &   0.03 &   37.02 &   38.58  &  -1.56  &    \\
   4 &  00:54:50.16 &  -37:35:55.14  &    5.89  &  57.10 &  \ldots &   0.02 &   37.01 &   39.05  &  -2.04  &    \\
   5 &  00:54:46.50 &  -37:36:32.76  &    5.89  &  57.10 &  \ldots &   0.21 &   37.33 &   38.97  &  -1.63  &  H \\
   6 &  00:54:44.44 &  -37:36:40.24  &    5.89  &  57.10 &  \ldots &   0.16 &   36.99 &   38.76  &  -1.77  &  H \\
   7 &  00:54:54.67 &  -37:36:41.15  &    5.89  &  57.10 &  \ldots &   0.17 &   37.30 &   38.75  &  -1.46  &  H \\
\ldots \\

  % 2 &  00:54:44.00 &  -37:35:18.69  &   2.94  &  28.55  &  101  &  37.27  &  36.13  &  38.54  &  37.57  &  37.62  &  37.42  &   0.41 &   37.36 &   38.71  &  -1.35  &  H \\
  % 3 &  00:54:41.38 &  -37:35:39.49  &   2.94  &  28.55  &   67  &  37.01  &  35.88  &  38.57  &  37.59  &  36.36  &  36.85  &   0.03 &   37.02 &   38.58  &  -1.56  &    \\
  % 4 &  00:54:50.16 &  -37:35:55.14  &   2.94  &  28.55  &   56  &  37.00  &  35.86  &  39.04  &  37.96  &  36.76  &  36.20  &   0.02 &   37.01 &   39.05  &  -2.04  &    \\
  % 5 &  00:54:46.50 &  -37:36:32.76  &   2.94  &  28.55  &  166  &  37.29  &  36.14  &  38.89  &  37.83  &  37.62  &  37.42  &   0.21 &   37.33 &   38.97  &  -1.63  &  H \\
  % 6 &  00:54:44.44 &  -37:36:40.24  &   2.94  &  28.55  &   65  &  36.96  &  35.81  &  38.70  &  37.69  &  37.32  &  37.27  &   0.16 &   36.99 &   38.76  &  -1.77  &  H \\
  % 7 &  00:54:54.67 &  -37:36:41.15  &   2.94  &  28.55  &  162  &  37.26  &  36.11  &  38.68  &  37.68  &  37.32  &  37.27  &   0.17 &   37.30 &   38.75  &  -1.46  &  H \\
						
\hline

\end{tabular}
\end{minipage}

\caption{\rm{Catalogue of regions identified in NGC 300. Full catalogues for all the galaxies are available in the online version of the paper. Region ID is not continuous since regions rejected based on their location within the image are assigned IDs as well. The aperture diameters ($d_{\rm ap}$) are quoted both in angular and physical scales, using the distances listed in Table \ref{data_sample}.  $A_{\rm FUV}$ is the attenuation in the FUV band derived using the \citet{Hao2011} relationship. The following columns list dust corrected  $\log(L_{\rm H\alpha})$ and $\log(L_{\rm FUV})$ values. $Y_{\rm corr}=\log(\frac{L_{\rm H\alpha, corr}}{L_{\rm FUV, corr}})$ is the logarithm of the regions' \HFratio flux ratio after the dust attenuation. The regions marked with `H' in the final column are in the hand-picked clean sample. Six columns have been omitted in the printed version for compactness. These contain $\rm H\alpha$, FUV and MIPS $\rm{24 \mu m}$ luminosities and the associated photometric errors. }}
\end{table*}

In all the figures presented in this section of the paper, the error bars associated with the data points correspond solely to the propagation of errors associated with the photometric measurements and the uncertainty in the  measurement of each galaxy's distance as quoted by the source.
Systematics originating from physical misalignment of emission peaks in different wavelengths and variation in dust attenuation corrections are discussed towards the end of this section.

\subsection{Full sample} \label{results}

\begin{figure*}
\begin{minipage}{180mm}
\centering
\includegraphics[width = 1.0\columnwidth]{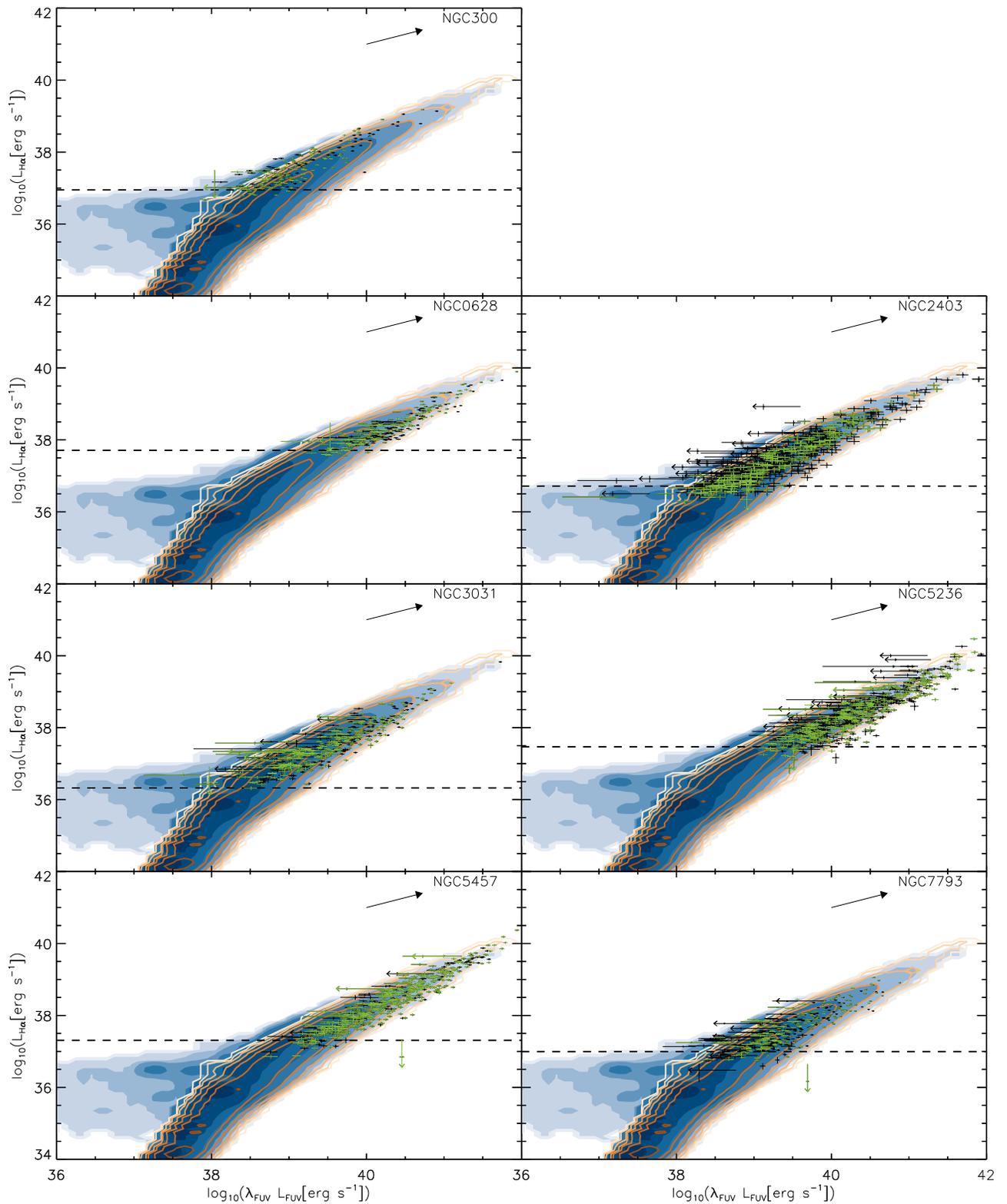}
\end{minipage}
\caption[Results: rest of the sample]{\Ha vs FUV luminosity plots of \Ha selected region population. Each panel presents the results for one galaxy.  Dust attenuation corrections have been applied.  Black arrows represent a dust attenuation correction vector corresponding to $A_{\rm H\alpha}=1 mag$. \rm{Blue filled contours are PSS models,  orange contours are CMDMSM models, as in left column of Figure \ref{fig:models}.} The $50$ per cent completeness limits are displayed with dashed lines.  Light green points represent the manually verified sample of clean regions. Black points are the remaining identified objects. }
\label{7gals_HvF}

\end{figure*}

Figure \ref{7gals_HvF}  presents the behaviour of the \Ha selected region sample in each of the investigated galaxies. 
The data points have been corrected for the effects on the dust attenuation using Equation \eqref{afuv}. 
We chose to plot \Ha luminosity rather than \HFratio in order to illustrate more clearly the behaviour of regions with flux limits rather than confident flux measurements.

Note that the diagnostic wing of high \HFratio ratio models in the PSS population is now represented as the broad horizontal band in \Ha.
The high luminosity trend for both models is equivalent to a constant \HFratio ratio illustrated in \rm{Figure \ref{fig:models}.}
The turnover in the CMDMSM models is less pronounced in this parameter space.  
Note that the attenuation vector is at a shallower inclination than the high luminosity trend. 

The green points represent a manually verified subsample of the \textsc{SExtractor} detections, referred to later as the ``clean'' sample.  
It was obtained in the following manner.  First, all regions were visually checked against the \Ha native resolution image.  
We removed from the sample the regions excessively affected by crowding or deemed to be a part of a larger scale structure.
The majority of regions eliminated at this step were connected with \HII regions displaying shell morphology.  

The second step was the removal of regions without a clear match in the FUV image or with a match that was significantly spatially offset.
This was to eliminate unreliable FUV flux estimates.
In cases where several distinctive \HII regions were  associated with the same large FUV structure, they were rejected unless a matching FUV substructure was seen.
Regions with no associated FUV peak that were located a large distance from any prominent FUV features were retained in the sample.

The population of identified regions follows the theoretical distribution closely in the high luminosity regime. 
There is a \rm{small number} of regions present above the main trend. A number of those objects have upper limits on FUV luminosity, which is consistent with their expected low FUV fluxes.  Regions with \lfuv \ upper limits at luminosities exceeding $10^{39} \rm erg\ s^{-1}$ \rm{typically} have low photon counts and high dust attenuation corrections. 
The clean sample displays an even tighter correlation with the predicted trends. 
\rm{The most FUV luminous regions have somewhat lower \HFratio ratio values than those predicted by the models. This could indicate a presence of older stellar population within the aperture. This effect could also be introduced by overcompensating for dust attenuation effects (see Section \ref{dust} for more detailed discussion). }

We note that adoption of \Ha as the reference wavelength means we are not sensitive to tracing the turnover in the \HFratio flux ratio.  Indeed, none of the galaxies present evidence of this effect within the \Ha selected population. 
As expected, the turnover is manifested in the FUV selected populations, which are not presented \rm{in this paper}.

In the low luminosity end we see a small number of regions that begin to deviate from the CMDMSM prediction while remaining in agreement with the PSS models.  
This populations is \rm{found within NGC 300, NGC 2403, NGC 3031 and NGC 7793.}  This is expected, since these galaxies are the least distant ones within our sample.  Therefore, they offer an enhanced sensitivity towards the identification of low luminosity objects.

\begin{figure*}
\begin{minipage}{180mm}
\centering

\includegraphics[width = 1.0\columnwidth]{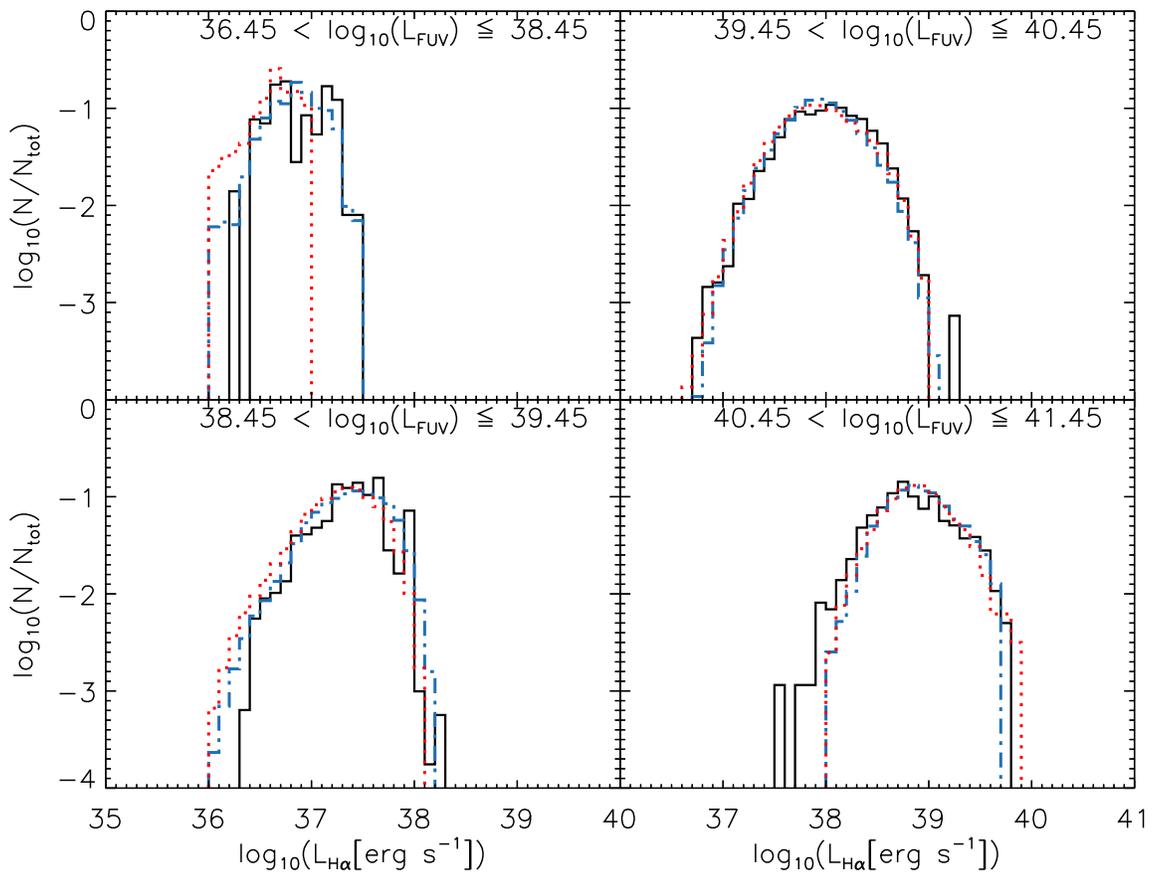}
\end{minipage}
\caption[\Ha luminosity stack]{\Ha luminosity histograms for objects in a given FUV luminosity bin.  Blue dot-dashed lines represent the PSS model population.  Red dotted lines represent the CMDMSM model population. Black solid lines represent the \Ha selected objects stacked from the sample of eight galaxies as displayed in Figure \ref{7gals_HvF}.  All luminosities are expressed in \ergps. Regions with flux limits have been excluded from the data histogram. Details of stacking are explained in Section \ref{results}}
\label{Ha_hist_stack}

\end{figure*}

\begin{figure*}
\begin{minipage}{180mm}
\centering
\includegraphics[width = 1.0\columnwidth]{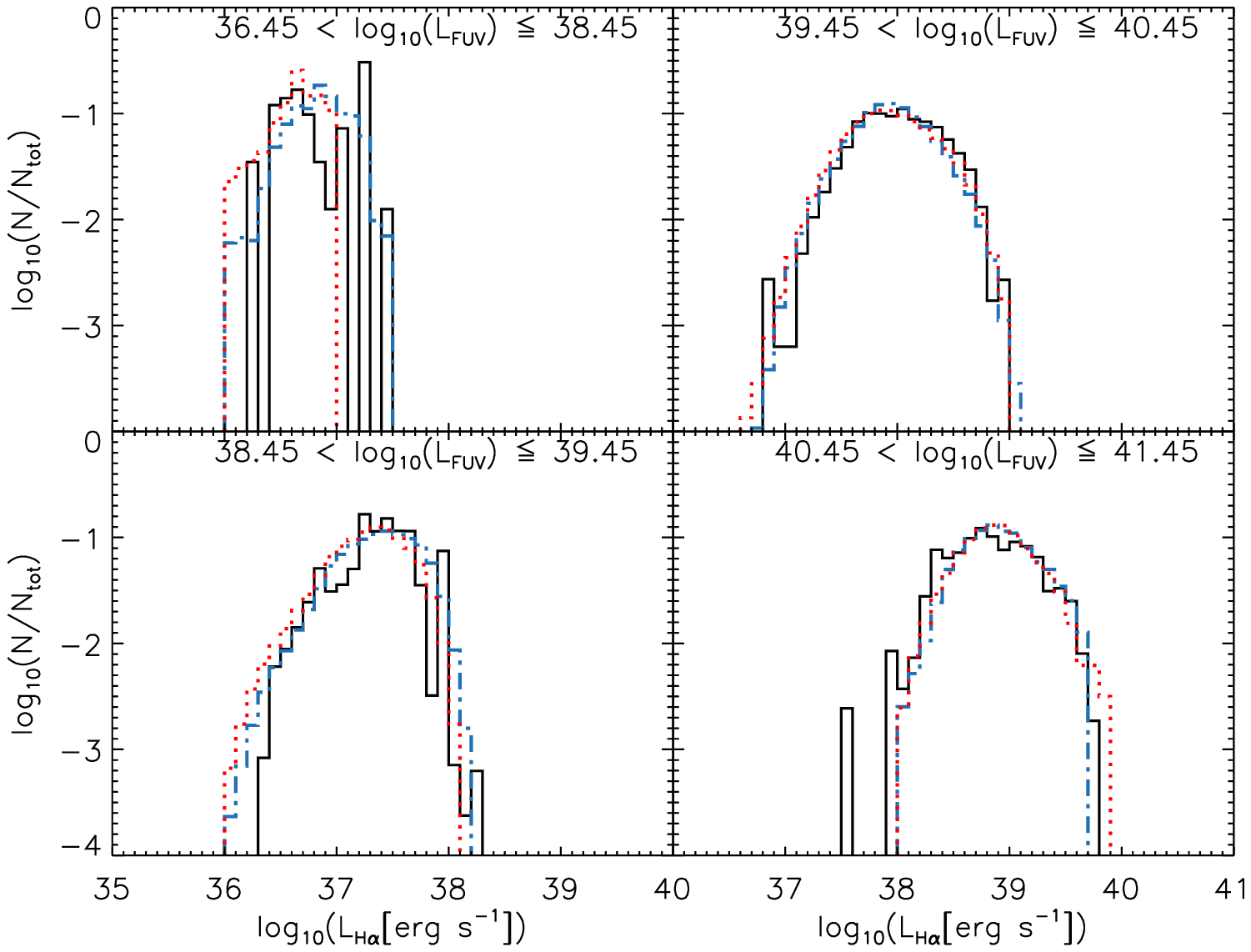}
\end{minipage}
\caption[\Ha luminosity stack - handpicked]{\Ha luminosity histograms for objects in a given FUV luminosity bin.  Histogram colours as in Figure \ref{Ha_hist_stack}. Data histogram now only includes \Ha selected objects from the clean subsample. As before, data are stacked from the sample of eight galaxies. }
\label{Ha_hist_stack_reghand}

\end{figure*}

Figure \ref{Ha_hist_stack} illustrates an attempt at statistical description of this effect.  Model and data populations have been divided into four luminosity bins and normalized histograms illustrating the distribution of  \Ha luminosity (\Ha luminosity functions, effectively) were made. Stacking has been performed by adding the normalized histograms for each $L_{\rm FUV}$ bin, weighted by the total \Ha luminosity of each galaxy as listed in Table \ref{data_sample}.  Each histogram was then renormalised.  Note that the histograms of data population do not take into account any of the regions with flux upper limits.

Given the behaviour of CMDMSM and PSS model contours, we do not expect to be able to distinguish between the models in the high-mass (i.e. FUV-luminous) regime.  Indeed, the histograms of model populations for the \rm{three} bins with $\log_{10}(L_{\rm FUV}[\rm erg\ s^{-1}]) > 38.45$ \rm{are} very similar. The models span the same ranges within \Ha luminosities and the detailed shapes of the normalised distributions are consistent with each other. The  $38.45 < \log_{10}(L_{\rm FUV}[\rm erg\ s^{-1}]) \leq 39.45$ bin offers a minute hint of an offset between the peaks of the distributions of PSS and CMDMSM models, but the diagnostic significance of this is minuscule.

This changes as we investigate the \rm{lowest FUV luminosity bin}.  The $36.45 <  \log_{10}(L_{\rm FUV}[\rm erg\ s^{-1}]) \leq 38.45$ \rm{bin encompasses the luminosity range spanned by the diagnostic high \HFratio ratio wing.  We see a disparity between the peaks of the two model distributions. 
Note also that within this bin the PSS models cover an \Ha luminosity range extending to values higher than those predicted by CMDMSM regions by $0.5$dex.}

Figure \ref{Ha_hist_stack} illustrates that the data population is  consistent with the models within the highest luminosity bins.  The luminous outliers seen in the highest \lfuv bin can be readily explained when looking at pre-binned Figure \ref{7gals_HvF}.  
They correspond to the luminous regions within NGC 5457 that lie on the extension of the constant \HFratio ratio trend.  Our theoretical models simply do not extend to high enough masses to account for them.  The low luminosity outliers within \lfuv bins are adressed in a following section. 

A key result is the presence of observed regions above the maximum \Ha luminosity cutoff predicted by the CMDMSM models.  The effect is clearly seen in the two lower \lfuv bins.   Those regions fall within the \Ha luminosity range predicted by PSS models.  

Figure \ref{Ha_hist_stack_reghand} is constructed in a manner analogous to Figure \ref{Ha_hist_stack}, by considering only the manually verified clean subsample.  Overall, the clean subsample reduces the number of available regions by $\approx 50$ per cent. This becomes most significant within the lowest \lfuv bin, with the population within the cut reduced from $40$ to $16$ objects. 
Despite the reduction in size, the clean sample  still exceeds the CMDMSM predicted \Ha luminosity cutoffs within the two lowest \lfuv bins.  The consistency with both model populations in the high \lfuv regime is also maintained.

\subsection{Effects of region selection}

A major challenge in investigating resolved populations of star forming regions is identification of the regions of interest.  The approach described in Section \ref{section:detection_method} has been selected to maximize the identification efficiency.  However, there are a few caveats associated with using automated region selection.  

Our analysis is performed on the star forming regions within the optically bright discs of the spiral galaxies.  
In this environment, detection completeness becomes a two-dimensional function of the position within the image due to the effects of crowding.

We have performed Monte Carlo tests to study the detection efficiencies within the galaxies.  
We have inserted artificial objects into the image of each galaxy and checked their recovery rate by the pipeline.
The test objects were point sources of various intensities convolved with the \textit{GALEX} FUV PSF.
We placed 20 objects at a time for 300 realisations per flux threshold tested.
\Ha fluxes investigated ranged from $10^{36.0}$ \ergps to $10^{39.8}$ \ergps in steps of $0.2$ dex.
The probability of  placing an artificial object at a particular location within the image has been weighted by the intensity of the smooth component of FUV emission.
In each case we applied our processing pipeline to the modified image.
A recovery was considered successful if a region was identified within 4 pixels ('set 2') or 2 pixels ('set 1') of the original placement with measured flux within $0.2$ dex of the input 

The $50$ per cent completeness thresholds plotted within Figure \ref{7gals_HvF} is the flux at which the successful recovery percentage in the Monte Carlo tests is $50$ per cent. This value is extrapolated between the \Ha luminosity bins. 
We find the detection threshold within the galaxy to be typically an order of magnitude higher than for objects placed outside the galactic disc.  
We also find that changing the desired flux recovery accuracy to $0.1$ dex does not typically increase the completeness threshold by more than $0.2$ dex.

\begin{figure*}
\begin{minipage}{180mm}
\centering
\includegraphics[width = 1.0\columnwidth]{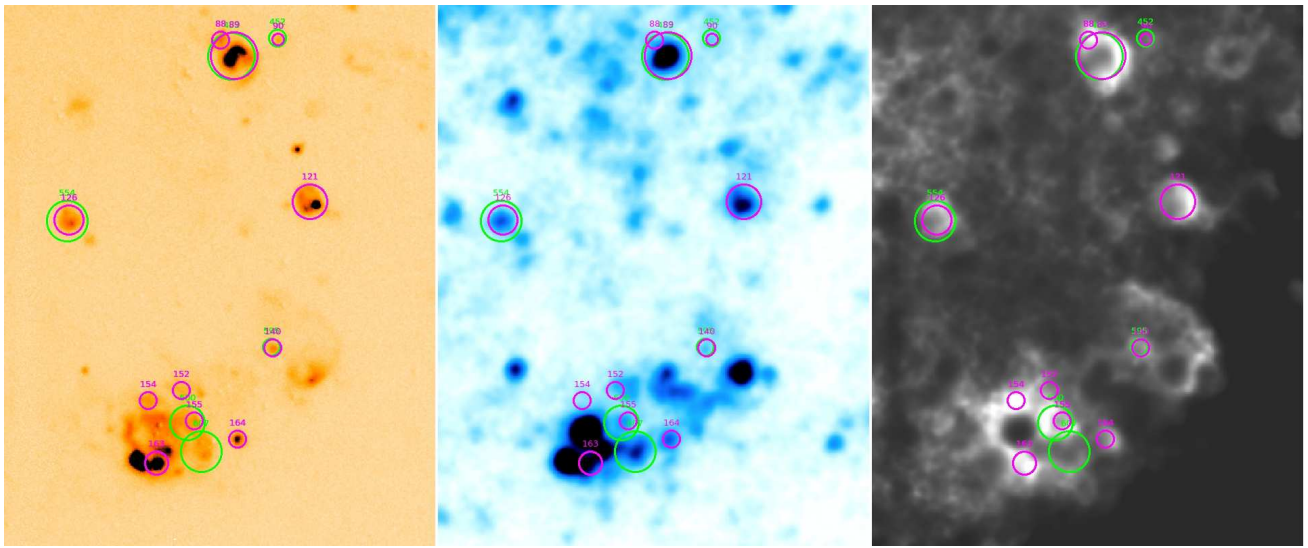}
\end{minipage}
\caption[NGC300: candidate regions]{A field within the SW part of NGC 300.  North is up, east is left. \textit{Left}: \Ha image with the smooth component subtracted, inverted colour scale. \textit{Middle}: FUV image with the smooth component subtracted, inverted colour scale. \textit{Right}: \HFratio ratio map of the two images, obtained by convolving the \Ha image to FUV resolution and then dividing it through by the FUV image.  Green circles are \Ha selected regions, magenta circles are FUV selected regions. Note that the \HFratio map in the rightmost panel is dominated by features originating in areas lacking any physical objects and so is not particularily useful as an identification tool. It is useful, however, as means of confirmation of our detections.}
\label{NGC300_tripleview}
\end{figure*}

An important issue related to crowding is the structural complexity of the star formation regions. 
As a rule, the automated object identification algorithms tend to split regions with complex morphologies into constituent substructures.
This issue tends to be even more severe in some of the other investigated object identification software, e.g. \textsc{HIIphot} \citep*{HIIphot}.
While \textsc{SExtractor} is less prone to splitting the complexes into subcomponents, the issue is still present in our catalogues, although to a lesser extent. 
Bear in mind, this is similar to the approach adopted in the manually compiled catalogues of \HII regions presented by \citet{HK83}, where most of the \Ha peaks are identified as individual regions.

\begin{figure}
\centering
\includegraphics[width = 1.0\columnwidth]{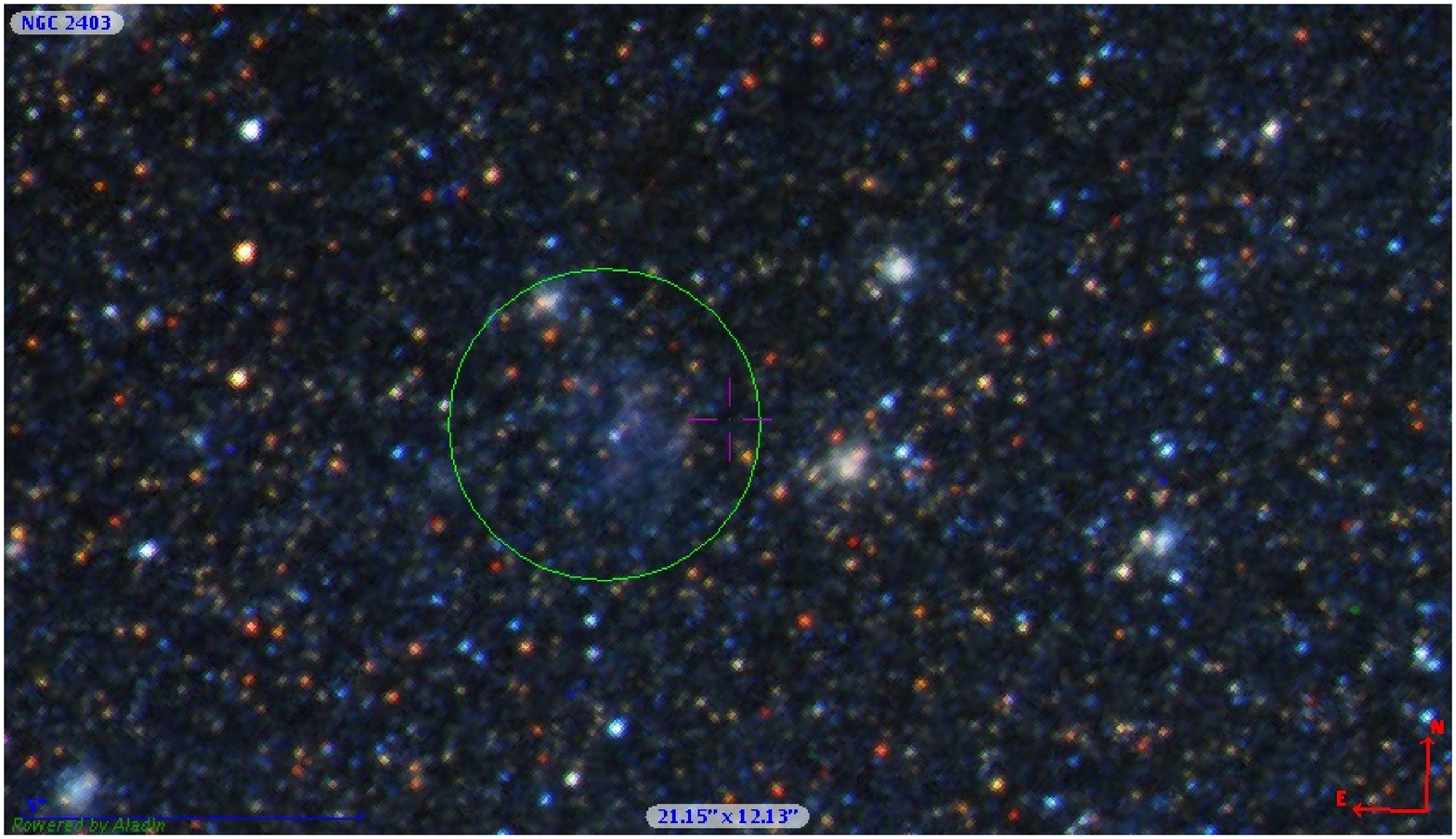}
\includegraphics[width = 1.0\columnwidth]{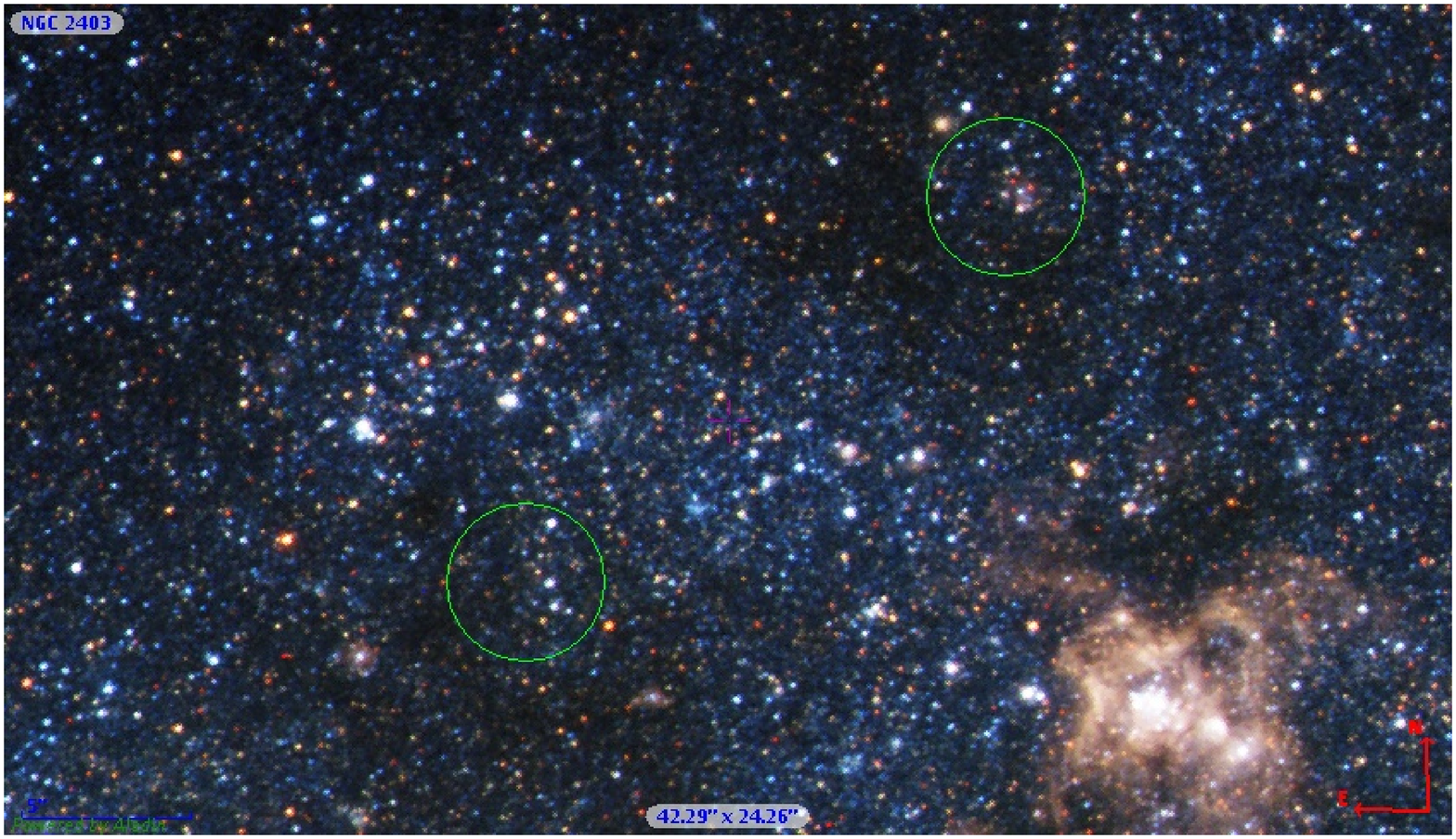}
\caption[NGC 2403: candidate regions HST]{HST view of two fields in NGC 2403, colours are B,V,I+\Ha.  The green circles are three objects exceeding the CMDMSM-predicted \HFratio flux ratio values.  The image was taken from the Hubble Legacy Archive. North is up and east is left. The top panel is approximately 21 by 12 arcsec, the bottom panel 42 by 22 arcesec. Images were produced using the \textsc{Aladin} software. }
\label{N2403:HST}
\end{figure}

Figure \ref{NGC300_tripleview} shows a field in the south-western part of NGC 300. 
Marked objects fall above the selection line illustrated in Figure \ref{fig:models}.
Objects from the \Ha selected sample are marked in green, the ones from FUV selected sample -- in magenta.
Note that a number of regions have been selected independently in both \Ha and FUV, with similar spatial extent.
The majority of objects exceeding the CMDMSM constraint are expected to be faint FUV sources.
Hence we do not expect to be able to identify them reliably using FUV selection.  
This is reflected by the dominance of \Ha selected objects in our candidate sample. 

The large star forming complex in the southern part of the field in Figure \ref{NGC300_tripleview} is a good illustration of the issues with region subdivision. 
We see a degree of substructure present both within \Ha and FUV data.
The components of the complex are treated as individual objects.
There are also spatial offset between the sources of ionizing radiation and the \Ha emission peaks within the comples. 
\Ha radiation has a clumpy shell structure, surrounding a pair of dominant FUV peaks.  
The shell morphology is clearly manifested in the \HFratio map.  
The large \HFratio values of those regions can therefore be explained in those cases by the lack of spatial coincidence between \Ha and FUV emission. 
This reiterates the importance of using a clean control sample in the study.

An additional qualitative verification with \textit{Hubble Space Telescope} (HST) broadband F606W and F555W images of the candidate regions was performed where the data was accesible via \textsc{Aladin} software \citep{Aladin}.  
This check reveals presence of star clusters at the majority of candidate locations. 
In some cases, three colour composites were available via the Hubble Legacy Archive.
Example morphologies of candidate regions are displayed in Figure \ref{N2403:HST}.

The object in the upper panel has what could be viewed as a model morphology, with a faint shell of \Ha surrounding a cluster with a luminous blue star present.  The objects in the bottom panel have been identified in a much more crowded area of the galaxy.  The northern region in that panel presents a clear cluster morphology. Interestingly, it contains bright red as well as blue, which could be evidence of multiple, spatially coincident star forming events.   
The southern object also contains luminous blue stars, but there is a chance of external ionizing flux contamination associated with neighbouring populations.

To recapitulate, the detections within the outskirts of large complexes can be unreliable, especially if wispy \Ha morphology is seen within the candidate aperture. 
In those cases the ionizing radiation originates outside the aperture containing the \Ha source. 
This leads to an artificial boost in the observed \HFratio ratio. 
Faint \Ha peaks identified in relative isolation are considerably more reliable, with coincident star clusters revealed in HST images.

\subsection{Dust attenuation consideration}\label{dust}

In this section we justify the necessity for correcting for dust attenuation. 
We also discuss the influence the choice of dust correction recipe has on the results of our study and the error balance.
 
First, let is consider the effect of the chosen correction method of \citet{Calzetti2007}.
An alternative approach using the relationship between the FUV and MIPS $\rm{24 \mu m}$ luminosities is also possible, as described by \citet{Hao2011}.  Finally, it is possible to determine the attenuation values independently of each other, which means direction of the attenuation vector no longer remains fixed by the extinction curve. 
Note that this approach presumes to allow for the variation in the relative geometry of dust, stars and ionized gas.

\begin{figure*}
\begin{minipage}{180mm}
\centering
\includegraphics[width = 1.0\columnwidth]{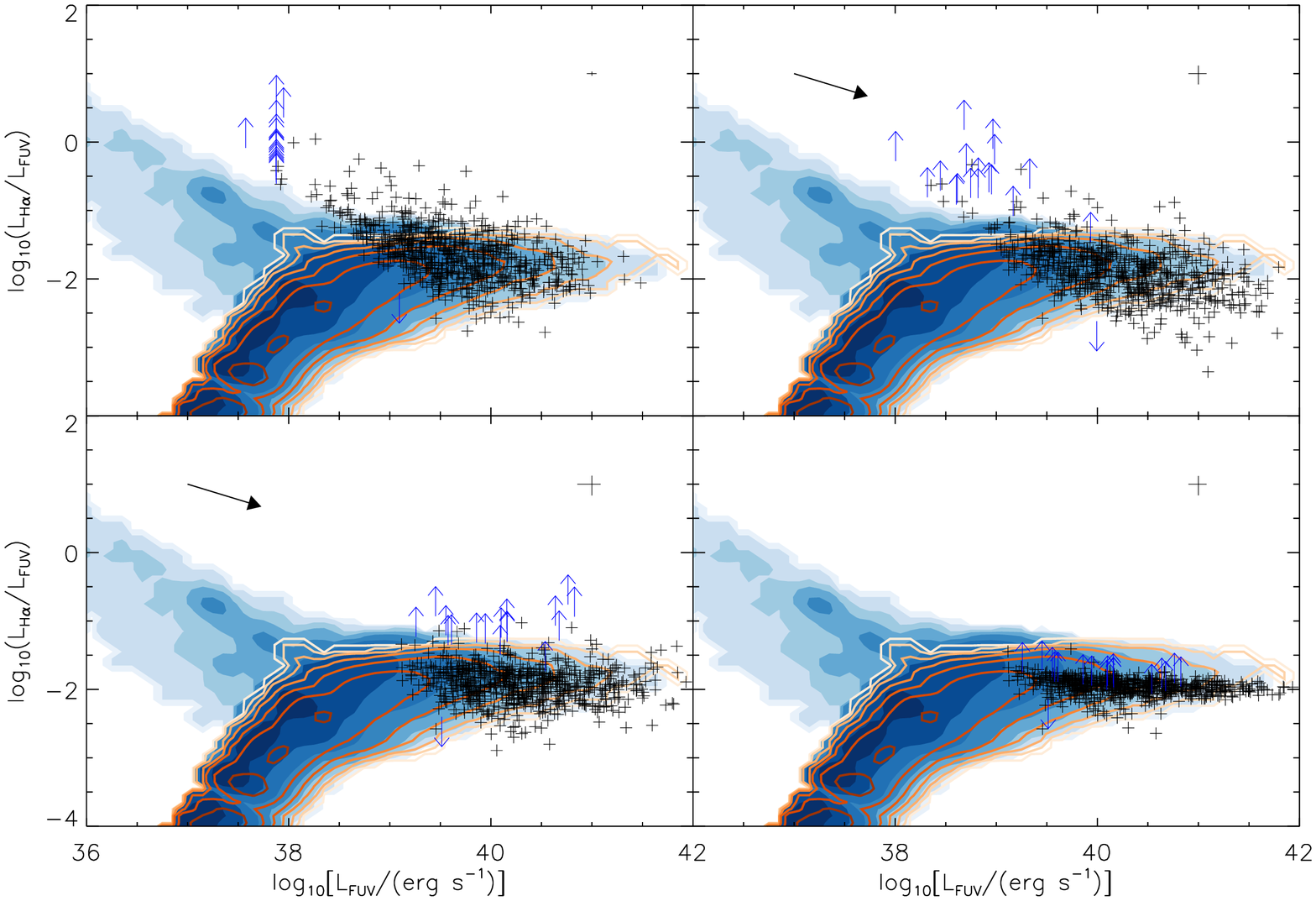}
\end{minipage}
\caption[Results: NGC 5236]{\HFratio flux ratio plots of \Ha selected region population in NGC 5236 plotted against FUV luminosity with different approaches to dust attenuation corrections.  \textit{Top left}: no correction applied. \textit{Top right}: both $A_{\rm FUV}$ and $A_{\rm H\alpha}$ derived from  \citet{Calzetti2007} relation. \textit{Bottom left}: both  $A_{\rm FUV}$ and $A_{\rm H\alpha}$ derived from \citet{Hao2011} relation. \textit{Bottom right}:  $A_{\rm FUV}$ and $A_{\rm H\alpha}$ derived independently.  Note that independent approach converges to a constant \HFratio ratio. Contours are the same as in Figure \ref{7gals_HvF}.  Black crosses represent individual data points with error bars ommitted for clarity.  Blue arrows represent limits on the ratios, where applicable. Lower limits on a flux ratio correspond to upper limits on \lfuv. Black arrows represent a dust attenuation correction vector corresponding to $A_{\rm H\alpha}=1\ \rm{mag}$.  \rm{The median errors are indicated in top right corner of each figure.}}
\label{attenuations}

\end{figure*}

Figure \ref{attenuations} illustrates the results of these three different approaches to the derivation of attenuation values.  
The data points are plotted with error bars omitted for clarity. 
\rm{Median error bars are indicated in each of the plots}. 
The points that are lower limits in \HFratio ratios are also upper limits in \lfuv.  
Top \rm{left panel shows the uncorrected fluxes for reference. Top right}panel illustrates the results of adoption of \citet{Calzetti2007} recipe, \rm{ while bottom left} panel shows the application of  \citet{Hao2011} relation. 
The bottom \rm{right} panel presents a combined  method where $A_{\rm H\alpha}$ is derived from the \citet{Calzetti2007} prescription, while $A_{\rm FUV}$ is derived from the \citet{Hao2011} relationship.

Prior to the application of any corrections, a significant number of regions are present at \HFratio values exceeding the theoretical predictions.
We note a group of objects within the locus of the diagnostic branch of PSS models. 
Accounting for dust attenuation effects is necessary in order to ensure the physicality of presented populations.

A striking feature of the plots is the convergence of data points in the combined method (bottom right panel) to a constant flux ratio value of $\log(L_{\rm H\alpha}/L_{\rm FUV}) \approx -2.0$. This is a direct result of the assumptions within the derivations of the two relations.  Both \citet{Calzetti2007} and  \citet{Hao2011} have calibrated their empirical relations against integrated properties of the galaxies.
 Therefore, the combined method recovers the intrinsic \HFratio ratio of the assumed underlying stellar population.
Note also that the reason why this approach converges to a value lower than the median of our high-mass models is because we are only investigating populations less than 8 Myr old. 
Including the older objects would reduce this value.

With the combined method excluded, we need to investigate the behaviour of \citet{Hao2011} and \citet{Calzetti2007} relations when applied to regions with  intrinsic \HFratio ratios deviating from the galactic average.
The fundamental issue is the dependence of the IR-derived dust attenuation corrections on the intrinsic \HFratio ratio of the objects.
For any region with a stellar population too old to produce significant ionizing flux (i.e. older than approximately 10 Myr in the classical view) the \citet{Calzetti2007} method will overestimate the effect of dust attenuation. 
This is because the age-induced intrinsic lack of \Ha flux will be wrongly attributed to dust absorption.
This effect is the origin of the high $\lambda_{\rm FUV}L_{\rm FUV}$ and low \HFratio value regions in the top right panel of Figure \ref{attenuations}.

The question becomes more complex in the case of  clusters with high \HFratio flux ratios.
Those objects normally do not populate the IMF fully so that they are intrinsically deficient in the FUV-generating B-type stars.
The FUV radiation field generated by those stars is normally the primary source of heating of the dust within the cluster site. 
Hence, any of the available calibrations becomes problematic. 
The changes in the thermal balance of dust due to missing B stars are expected to cause the \citet{Calzetti2007} calibration to underpredict attenuation values.
This could lead to false positives within the diagnostic regime. 

An additional layer of complexity is added by the effects of geometry. 
The galaxy-wide calibrations average over large numbers of lines of sight within the disc, making simple geometry models more applicable. 
Here we are considering individual objects with \HFratio balance dominated by only a few stars.
The relative distributions of the dust grains, the ionized gas and the stars themselves become of a key importance. 
Exploring these issues would require a complex suite of radiative transfer models \citep[e.g. with the \textsc{Cloudy} code as described by][]{Cloudy}, which remains beyond the scope of this study.

A variation in geometry corresponds to a change in the effective attenuation curve, modyfing the proportionality between $A_{\rm H\alpha}$ and $A_{\rm FUV}$. 
We are fully aware of the associated caveats. 
As demonstrated earlier, however, independent attenuation corrections only recover their implicit assumptions.
We have no feasible means of infering anything about the line of sight towards the individual clusters.
Therefore, we apply a straightforward extinction recipe to achieve a best average performance across our sample.

To avoid this issue, we adopt the recipe of \citet{Hao2011}.
Although the use of an \Ha based recipe seems intuitive given that we are using \Ha as reference wavelength for region identification, FUV-based calibration has number of advantages.
First, the question of dust thermal balance should be better adressed by the method tied directly to the heating radiation field. 
Second, it addresses the problem of missing FUV flux due to the use of \Ha based apertures.
The missing flux will increase the attenuation value.
This will then remove the region from the diagnostic locus, reducing the chance of false positives.

We have also tested the use of an alternative FUV based calibration of \citet{Liu2011} (equation 5). 
This recipe follows the same functional form as that of \citet{Hao2011}, with the coefficient on the MIPS term changed from $3.89$ to $6.0$. 
It was derived for younger stellar populations.  
We found that the higher attenuation values do not alter the location of most data points significantly.  The most significant effect is the reduction of the number of regions populating the diagnostic regime. 
We also observe the highest \lfuv data points forming a trend aligning with the direction of the attenuation vector, which is an indication of excessive attenuations.

\subsection{Discussion of uncertainties}

The error bars presented in Figure \ref{7gals_HvF} indicate the uncertainties associated with photometric measurements and the uncertainty in distance affecting the conversion from fluxes to luminosities.  Other sources of uncertainty are the flux calibrations themselves.  \citet{Kennicutt2008} quote the median uncertainties in their \Ha flux calibrations as $12$ per cent ($\sim 0.05$ dex). \textit{GALEX} calibrations \citep{GALEX_calibration} are expected to be accurate to within 1 per cent ($\sim 0.004$  dex), which is negligible when compared to the Poisson uncertainties and possible systematic effects. 

There are two potential sources of systematic errors in this study - the dust attenuation corrections and the possibility of inadequate selection of apertures. 
In the case of dust, we have selected the methodology so as to minimise the chance of systematics causing false positives within the diagnostic region. 
We appreciate that the geometry variations within the regions can plausibly alter the derived attenuation values by more than a magnitude.

The influence of aperture selection is also hard to quantify. Spatial offsets between FUV and \Ha peaks have been observed before \citep[e.g.][]{Calzetti2005}. Crowding increases the challenge of assigning the correct fluxes to individual regions. The visually selected clean sample has been created to address these issues and has shown results consistent with the results from the full data set.  Validation of candidate regions against HST photometry has also ensured the observed diagnostic subpopulation is not merely an effect of a systematic error.
Furthermore, any objects where only flux limits are available have been excluded from the histograms in Figures \ref{Ha_hist_stack} and \ref{Ha_hist_stack_reghand}.
Nevertheless, we acknowledge that the systematics originating from the source matching and background subtraction can modify the fluxes by factors up to $0.5$ dex in the most extreme cases.

\subsection{Statistical significance of the results}

\rm{Statistical treatment of this type of study is made complex by }the non-trivial \rm{variation of completeness limit with position, as well as} the intrinsic systematics due to the dust attenuation correction and our ignorance of any flux leakage. 
Given our simplistic galaxy-averaged treatment of the completeness effects \rm{and the lack of model free parameters}, we do not \rm{require} our models to \rm{return a nominal high probability in a single model} statistical test. \rm{Instead, we focus on the relative likelihoods of the models.}

\rm{The most general approach in this case is to} perform a simple Poisson-type test in the regime of our diagnostic high \HFratio ratio wing. 
We take the lowest \lfuv bin of Figure \ref{Ha_hist_stack} with the total population of $46$ objects and split it into two subgroups according to the dust-corrected \Ha luminosity. 
We observe 26 objects with $log_{10}(L_{H\alpha}/ {\rm erg\ s^{-1}}) \leq 36.8$  and 20 objects with $log_{10}(L_{H\alpha}/ {\rm erg\ s^{-1}}) > 36.8$.  
When considering the clean region sample, we observe 22 objects within the lowest \lfuv bin. those values are 13 and 9 respectively. We can see that the proportion of populations within these bins remains in excellent agreement.
The threshold \lha value has been selected to be close to the maximum value allowed for the CMDMSM models.  

We estimate the number counts predicted by our models by multiplying the normalized model distributions by the total observed number of points. However, we wish to stress that the model distributions are very sensitive to the accuracy of the approximate treatment of the completeness limits of our sample.  We note that the completeness corrections have been derived in a galaxy-averaged manner for dust-uncorrected region properties.  Therefore, we do not expect either of the region \rm{populations} to produce a \rm{nominal} high confidence fit to the observed data sample in terms of a strict hypothesis testing approach.  Despite quoting the probability values, we are most interested in the relative probabilities between the models.

PSS model populations predicts $N_{low,PSS}=37.8$ data points within the low $L_{H\alpha}$ bin and $N_{high,PSS}=8.1$ within the high $L_{H\alpha}$ bin.  
Assuming a Poisson distribution, this corresponds to the probability values of $P_{low,PSS}\approx 0.028$ and $P_{high,PSS} \approx 0.00034$ respectively.
CMDMSM models predict   $N_{low,CMDMSM}=45.0$ and $N_{high,CMDMSM}=0.9$, corresponding to $P_{low,CMDMSM}\approx0.0015$ and negligible $P_{high,CMDMSM}$ \rm{(nominal value of $\mathcal{O}(10^{-16})$).}
The combined formal likelihood ratio favours the PSS models by more than ten orders of magnitude. While the exact numerical values are \rm{subject to completeness effects}, a strong preference for the PSS models is made apparent. 

% $P_{high,CMDMSM}\approx1.11*10^{-16}$
\subsection{Notes on individual galaxies}

\subsubsection{NGC 300}

The \Ha data presented by \citet{Larsen1999} do not cover the whole extent of the optically bright disc of this nearby spiral.  Our calibration has been performed by matching fluxes of corresponding apertures on \citet{Larsen1999} and wider field, lower resolution image presented by \citet*{Hoopes_N300}.  
The scaling was then set to match the published flux of \citet{Kennicutt2008}.  
We acknowledge that this affects the quality of the calibration adversely and may contribute to the seen offset from the models. 
This offset could also be explained by the lower metallicity of NGC 300, as reported by \citet{Bresolin2009}. 
The minimum aperture diameter used was 15 pixels.  

% \subsubsection{NGC 628}

\subsubsection{NGC 2403}

As explored in Section \ref{section_sample}, this paper presents a new continuum-subtracted \Ha image of this galaxy, following an astrometry recalibration.  A large subsample of regions within the diagnostic regime were verified against HST images. This galaxy has larger than average uncertainties on individual regions' luminosities, arising primarily due to a significant uncertainty in the distance estimate, following the study of \citet{Freedman2001}.

% \subsubsection{NGC 5236}

 \subsubsection{NGC 3031 and NGC 5457}

The available \Ha observations for these galaxies taken from \citet{Hoopes2001} are of low resolution.  
This limits our ability to detect the regions that would fall within the diagnostic regime.   
The high dust content within those two targets magnifies the importance of an appropriate attenuation correction. 
Using the \citet{Calzetti2007} recipe leads to a much larger number of regions being identified within the diagnostic regime.

% \subsubsection{NGC 7793}

\section{Discussion} \label{discussion}

\subsection{Comparison to the BPASS models} %and their interpretation

We have demostrated earlier that the majority of regions within our sample are well matched to the model BPASS populations in the high luminosity regime.  
This is not surprising, since we expect \Ha selection to return mostly young star forming regions within the modelled age range of 1-8 Myr.  The regions at these luminosities have higher masses ($M_{\rm ecl} \ge 10^{3.5}M_{\rm \odot}$) and so we expect them to fully populate the IMF.  Stochastic effects become unimportant in this mass regime, as demonstrated previously by \citet{Koda2012}.

We also observe a small number of objects with \HFratio ratios lower than our predictions. These are seen clearly e.g. in figure \ref{attenuations}.
The objects in question can be explained by two effects. 
First, we are not incorporating the effects of flux leakage into our models. 
This is especially significant for the \Ha luminosities of our regions - in practice, the fraction of ionizing radiation escaping from the Str\"{o}mgren sphere volume into the surrounding space may exceed $50$ per cent \citep*[e.g.][]{OeyKennicutt,Zurita2001,Relano2002,Eldridge&Relano}. 
The leakage of ionizing radiation could explain the lowered \HFratio ratios of our objects \citep[see also][]{Relano2012}.

The second explanation is age of the regions. 
Given the 1-8 Myrs age range of our models, we expect clusters containing stars older than this upper age limit to have lower \HFratio ratio values. 
Star clusters with ages below 100 Myr are expected to retain significant \lfuv output after the massive O stars producing the majority of ionizing flux have completed their lifecycles.

The detection of this subpopulation is to be expected within our setup.  
Selection based on \Ha luminosity limits the number of regions detected at low \HFratio ratio values.  
Adopting the FUV selection reveals a much larger number of objects within this locus.
The full description of the FUV selected sample lies beyond the scope of this paper, since it offers no insight into the problem of the $m_{\rm max} - M_{\rm ecl}$ relation. 
This topic is going to be investigated in a future publication (Hermanowicz \& Kennicutt, \textit{in prep.}).

\subsubsection{Nature of high \HFratio ratio objects}

\rm{In this study we detect a subsample of star forming regions within the nearby galaxies that appear to be inconsistent with the CMDMSM models drawn using a truncated stellar IMF.
Our models suggest those objects are young stellar clusters of low masses ($M_{\rm ecl} < 10^{3.5}M_{\rm \odot}$), which nevertheless contain a massive WR or O type star with initial mass in excess of 20$M_{\odot}$. }

Here we explore the alternative explanations for the presence of the high \HFratio ratio population. 
First, these objects may all be claimed to be results of different systematics.  
The misalignment of \Ha and FUV peaks may indeed cause elevated \HFratio flux ratios in some areas, as we have discussed before.
Nevertheless, \rm{the clean region sample, compiled precisely in order to exclude systematics of this nature, displays the same departures from the CMDMSM predictions.} 

Alternatively, \Ha brightness peaks within the diffuse ISM could present as objects with high \HFratio values.
However, HST photometry reveals the isolated \Ha peaks to be typically associated with faint stellar clusters.
This proves that the $5\sigma$ selection cutoff is \rm{normally} sufficient to eliminate \Ha brigthness fluctuation artefacts from our sample.

A number of published studies report existence of massive stars within very low mass clusters.
\citet{Lamb2010} identified a number of candidate associations within the Magellanic Clouds dominated by a single O-type stars.  
Their estimates for the masses of these clusters and their dominant stars \rm{seem to be incompatible with the presence of a strong cluster mass dependent IMF truncation.} 
\rm{Furthermore}, \citet{Bressert&Bastian} recently presented a VLT-FLAMES Tarantula Survey results indicative of O-type stars forming in isolation or within very low mass associations.

\subsubsection{Potential for runaway star contamination}

Throughout the study we implicitly assume that any stars within the apertures have formed \textit{in situ}. 
However, our sample could be contaminated by high velocity runaway stars ejected from massive clusters.  
\HII regions associated with such runaways would have very little corresponding FUV emission and hence would naturally \rm{display high \HFratio flux ratios}.
We have no information on the kinematics of these objects, so we cannot exclude a runaway scenario directly.

Similar concerns have been raised regarding the stellar associations identified by \citet{Lamb2010}. 
\rm{The stellar velocity structure of those objects is unknown, it is therefore possible that their most massive stars are in fact runaways from larger clusters.  
The neighbouring low-mass stars may be merely at a coincident location and not a part of the same star formation event.}
\rm{A recent study by \citet{Gvaramadze2012} presented a bow shock detection around one of the stars in that sample as evidence of its runaway nature. 
The same study presents evidence supporting a runaway nature of the remaining stars in the \citet{Lamb2010} sample, based on spectroscopy of \citet{Massey2009}.}
\rm{In the case of the sample of \citet{Bressert&Bastian}}, the radial velocities of the target stars have been measured and the probability of the observed objects being line-of-sight runaways has been greatly reduced.  However, they could still be runaways within the plane of the sky or \rm{bound stars that are} close to the apocentre of a highly eliptical orbit. 

The velocity distribution of runaway stars is somewhat uncertain. Runaway status is often difficult to verify observationally \citep[see review by][]{Stone1991}, especially for lower ejection velocities. Works of \citet{Leonard1991} and  \citet{Silva&Napiwotzki2011} find OB runaway star velocities as large as 700 km s$^{-1}$. However, \citet*{Hoogerwerf2001} and \citet*{Tetzlaff2011} find that the majority of stars have runaway velocities slower than 100 km s$^{-1}$. Simulations of the runaway star population by \citet{Fujii2011} and \citet*{EldridgeLangerTout} also show that massive runaways are typically ejected with velocities significantly lower than 100 km s$^{-1}$ and that more massive stars attain lower velocities.  Therefore, the occurance of massive high velocity runaways remains possible, but is nevertheless rare.

\citet{Anderson2012_SNe} in a study of similar resolution to ours, found that 80 to 90 per cent of type Ib/c SNe occur in regions associated with H$\alpha$ emission in their host galaxies. The progenitors of those SNe are believed to be WR stars\citep*[e.g.][]{Podsiadlowski1992}. Furthermore, the distribution of the number of SNe closely follows the distribution of star-formation. The remaining 10 to 20 per cent occur in areas with no significant H$\alpha$ emission. The natural reduction in \Ha luminosity and cluster mass distribution implies that some SNe will occur in regions with H$\alpha$ emission below the detection limit. Therefore this study suggests that very few of the WR stars are runaways that travel a large distance from their birthplace.

Finally, it is uncertain whether any potential WR or O type runaways will produce observable \Ha emission. 
While every O or WR runaway produces ionizing photons, the density of the surrounding gas determines whether an \HII region is observed. 
Unless the local density is high, the \HII region might be extremely large and therefore of too low a surface brightness to be detected in our sample. 
If sources we have detected are associated with runaway stars, they must but be in a region of high ISM density so that the \Ha emission is constrained within a compact region less than $\sim70$pc in diameter.  
This corresponds to the ISM number density of the order of  $\sim1$ cm$^{-3}$ for O9 stars and $\sim10$ cm$^{-3}$ for O3 stars in the Str\"{o}mgren sphere approximation \citep{Osterbrock_book}. 
Those values are within the regime of giant molecular clouds \citep{Heiner2008} even prior to accounting for geometry effects and the ambient radiation field.    
Objects like this are known \citep*[e.g. $\zeta$ Ophiuchii, see][and the references therein]{Gvaramadze2012_ZetaOph}, but expected to be rare.  
Furthermore, assuming the proposition of \citet{Gvaramadze2012} that all the \citet{Lamb2010} O type stars are runaways, we note that only two of the stars have surrounding \Ha emission that could be discernible at the distances of the galaxies analysed in the paper \citep[based on the Magellanic Cloud Emission Line Survey images][]{Smith2000_MCELS}.

\rm{We also note that under the scale-free star formation paradigm proposed by \citet{Kruijssen2012}, most star formation in spiral galaxies is expected to be in unbound associations.  
Those slowly dissolving structures do not undergo dynamical relaxation and are not expected to produce high velocity runaways via dynamical interactions or violent gas expulsion. }

We expect objects populating our high \HFratio ratio branch to be of similar nature to the stellar associations identified in the studies of \citet{Bressert&Bastian} and \citet{Lamb2010}. Massive runaway stars are a potential contaminant, but considering all the factors discussed above, we do not believe that they account for all the regions in the diagnostic wing. 

%COMPARISON TO OTHER WORK FROM INTRO BEGINS HERE 

\subsection{Stochasticity and the truncation of the IMF}

This work is complementary to a number of previous studies that attempted to verify the \rm{presence of a cluster mass dependent truncation in the stellar IMF}.
\citet{Lee2009} attempted to observe the \HFratio ratio downturn proposed by \citet{Pflamm09} in the integrated fluxes of galaxies in the local Universe.
They found a small decline in the \HFratio ratio at low \lfuv values, in agreement with the semi-analytic IGIMF predictions.
\rm{However, they found the same decline to be consistent with models of \citet{Tremonti2007}, which assumed stochastic sampling. 
\citet{Fumagalli2011} extended this analysis with the use of their \textsc{SLUG} stochastic population synthesis code.  
They have demonstrated that the truncated IMF models significantly underpredict the integrated \Ha luminosities of galaxies in the lowest FUV luminosity bins. 
Finally, \citet{Weisz2012} extended the analysis of \citet{Lee2009} by including the estimated masses of the individual galaxies into their analysis, finding that implementation of truncated IMF significantly underpredicts the \HFratio ratios of the low-mass end of their sample. 
They offer time variations in the galactic star formation rate as an alternative explanation of the observed sample behaviour. 
We note their study does not account for the effects of stochasticity, though.}

A few studies similar to our project have been performed previously.
\citet{Corbelli2009} presented a study of star clusters in M33, focusing on the relation between the \Ha and bolometric luminosities. 
Their study revealed existence of clusters \rm{with \Ha fluxes in excess of what could be explained by models} with a \rm{cluster mass dependent} IMF truncation imposed. 
Our study presents evidence of presence of analogous objects in a number of more distant spiral galaxies. 
We also base our conclusions on an updated set of models, which include the effects of stellar evolution. 

\citet{Calzetti2010} used the HST to search for such objects within NGC 5194.  
Their investigation revealed a population of clusters (with $3\sigma$ detections in ${\rm H\alpha}$) consistent with fully stochastically sampled IMF, with no evidence supporting a presence of \rm{a cluster mass dependent IMF truncation}.  The results become less conclusive when the full object sample is taken into account. 
The study is based on a HST BVI-selected cluster sample described in detail by \citet{Chandar2011}. 
\citet{Koda2012} have recently presented a study of star formation in the XUV disc of M83 (NGC 5236), demonstrating their results to be consistent with stochastically sampled Salpeter IMF.  
We find our conclusions agree with the results of both those studies.  
\citet{Calzetti2010} focused on comparing the \Ha luminosity of the clusters to masses extrapolated from $U,B,V,I$ photometry.  
Our approach incorporates the directly observable FUV luminosities, hence directly \rm{addressing} the \citet{Pflamm09} predictions in the regime of individual clusters. 
\citet{Koda2012} focused on the number counts of \Ha- and FUV-bright clusters with a simple ageing relation explaining their relative number counts.
Their investigation of \HFratio flux ratios was limited to the integrated properties of the XUV disc, being parallel to the study of \citet{Lee2009}.

\rm{In all of those cases, our study complements the results by expanding the investigated sample of regions and allowing for attenuation variation between the individual objects. 
We also base our conclusions on an updated set of models, which include the effects of binary interactions on stellar evolution. 
Our study finds apparent disagreement with the \rm{truncated IMF model} predictions within the regime of individual clusters' properties, matching the conclusions of \citet{Fumagalli2011}, and \citet{Eldridge2012}. 
However, the observational constraints from the dataset presented in this paper are not sufficient to exclude weaker manifestations of the IMF truncation.}

\rm{Note that our data are not sensitive to differentiate between the random sampling preferred by \citet{Elmegreen2006} and the sorted sampling preferred by \citet{Weidner&Kroupa2006}.  
We note those two methods predict very similar property ranges for individual clusters.  
We also wish to stress that the work of PA07 and \citet{Pflamm09} are based on semi-analytic approximations and therefore unsuitable for stellar population synthesis using discrete stellar models. The PA07 relationship was derived to represent the typical $m_{\rm max}$ at a given $M_{\rm ecl}$ and stochastic departures from this relationship are expected, both towards higher and lower stellar masses. }

\rm{In line with the work of \citet{Cervino2003_AAP},} we conclude that the use of a fully stochastic stellar population synthesis code is critical in any study of small mass clusters and low SFR galaxies or environments. Codes like the widely used \textsc{Starburst99} \citep{Starburst99} are inherently unsuitable for the investigation of the effects of stochasticity.  
Fixed implementation of the IMF means that these codes are unable to represent the diversity of properties displayed by low mass clusters and young galaxies with low SFRs.
\rm{We suggest that stochastic codes are used in modelling of stellar populations in such cases.}

\subsection{Caveats}

There is a number of effects that are not accounted for in this study.  This section is a short overview of their relative importance.

Our analysis makes no provisions for the dynamics of the systems in questions.  One such a scenario could be a formation of a close binary in the centre of a low mass cluster consistent with the \rm{IMF truncation}, ejection of a significant portion of the cluster mass (probably mainly via ejection of the other massive stars) and then formation of a massive star with $m > m_{\rm max}$ via a coalescent binary scenario. 
Whilst we allow for effects such as common envelope evolution and mergers within the scope of our binary models, multi-star dynamical interactions are not included in the models. 
In practice, the stellar merger rate could be significantly boosted due to dynamical interactions within the cluster.  
Hence, we would see more CMDMSM-consistent clusters populating the diagnostic high \HFratio ratio area in Figure \ref{fig:models}.  
This would require an increase of the merger rate by two order of magnitude.

However, the recent numerical simulations presented by \citet{Oh&Kroupa2012} suggest that the effects of dynamical evolution on the $m_{\rm max}$ vs. $M_{\rm ecl}$ relation is negligible.
They find that only a small fraction of low mass embedded clusters will have the most massive star's mass significantly altered by stellar collisions.
The quoted study only considered clusters formed under IGIMF paradigm.
In view of these results, we do not expect the dynamical evolution to be responsible for the presence of high \HFratio ratio population in our data. 

\rm{A related concern is the potential effects of violent gas expulsion on the evolution of the observed clusters. Numerical studies  \citep[e.g.][and the references therein]{Baumgardt&Kroupa2007, Smith2013} suggest that it is not uncommon for the clusters to lose even $90\%$ of their associated gas mass and  as a result a significant population of stars formed in the original star formation event can become unbound. 
\citet{Smith2013} point out that the associated dynamical evolution of young clusters is a stochastic process highly sensitive to the initial conditions. 
However, under such a scenario we would expect the majority of the unbound stars not to have travelled far enough from our clusters so as to fall outside the apertures used in this study. }

The \textsc{BPASS} models do not include the effects of as rotation in stars, X-ray binary systems, or wind-wind collisions within the binary systems. 
The inclusion of any of these would lead to an increase in the ionizing radiation relative to FUV. 
In fact, recent models by \citet{Eldridge&Stanway2012} indicate that incorporation of rotation into massive binary star models lead to a 30 per cent increase in the ionizing flux. 

We are also assuming that the whole extent of the FUV radiation is originating solely from the underlying stellar population, hence ignoring any potential contributions from the ionized gas itself. 
Even though ignoring this contribution would artificially boost the \HFratio values of our objects, work by \citet{Danforth2003} indicates that even in the presence of shocked gas within a supernova remnant, the UV spectral energy distribution (SED) is clearly dominated by the stellar continuum.

In our analysis we do not consider the variations in the metallicity.  We base our analysis on the models of Solar metallicity, where in reality the metallicities vary from one galaxy to another \citep{Moustakas2010}.  Additional radial trends have also been observed within the discs of spiral galaxies \citep[e.g.][and the references therein]{Bresolin2009}.
Due to the issues with crowding, we rarely find any high \HFratio ratio regions close to the centres of studied galaxies.  
However, where the crowding becomes less of an issue, we do not observe a clear preference for identification of high \HFratio ratio regions at high galactocentric radii. 
The image of NGC 300 used in our study only covers the inner part of the galactic disk, and yet objects inconsistent with CMDMSM are identified there.

\section{Summary and conclusions}

We detect a small subpopulation of regions with high \HFratio flux ratios.
Comparison with model populations suggests those regions to be consistent with the results of stochastic sampling of an universal IMF, as suggested by \citet{Elmegreen2006}. 
\rm{We note that sorted sampling and PSS model populations are not easily distinguishable in the type of study presented here.}

We find that any study of those objects within the galactic discs is sensitive to the adopted dust attenuation.
An FUV-based recipe of \citet{Hao2011} was found to be most unbiased when combined with the effects of stochastically populated mass functions within the clusters. 
The high \HFratio ratio population persists after an application of dust attenuation corrections.  

The candidate region population is tested directly against the predictions of truncated IMF cluster models. 
We identify a subsample of objects whose properties seem to be inconsistent with \rm{a cluster mass dependent IMF truncation in the form described in Section \ref{section:models}}.
This view is consistent with a number of previous studies \citep{Corbelli2009,Calzetti2010,Weisz2012,Koda2012}.  

%However, the $m_{\rm max}$ vs. $M_{\rm ecl}$ relationships published up to now are not expected to accurately reflect any potential physical constraints on the star formation processes in individual clusters.

We emphasise the significance of stochastic effects in the studies of low SFR environments and caution against the application of semi-analytic recipes and classical stellar population synthesis code in this regime.

One of the factors limiting the scope of this work is the low resolution of GALEX data available.  Analogous studies will become more definitive with the future launch of World Space Observatory-UV planned for 2016 \citep{WSO}.

\section*{Acknowledgments}

We thank the anonymous referee for the comments leading to an improvement of this paper.
We thank Ben Davies, Mark Gieles, Mike Irwin, Diederik Kruijssen and Richard McMahon for useful discussions.
MTH thanks the STFC for his studentship.
This research has made use of the NASA/IPAC Extragalactic Database (NED) which is operated by the Jet Propulsion Laboratory, California Institute of Technology, under contract with the National Aeronautics and Space Administration (NASA).

\bibliography{bibliography}

\appendix
\section{Results for Calzetti attenuation correction}

\rm{ In this appendix we present alternative versions of the Figures \ref{7gals_HvF}., \ref{Ha_hist_stack}. and  \ref{Ha_hist_stack_reghand}, which have been created with the application of the \citet{Calzetti2007} dust attenuation relation, as opposed to the \citet{Hao2011} relation used in the main analysis.  
Note the higher scatter of objects around the model predictions, especially in the high luminosity regime. A larger number of objects is also observed in the diagnostic wing, making the constraint against the truncated IMF stronger.  This is because the \citet{Calzetti2007} attenuation relation is not as heavily biased against the regions with boosted \HFratio ratio.}

\begin{figure*}
\begin{minipage}{180mm}
\centering
\includegraphics[width = 1.0\columnwidth]{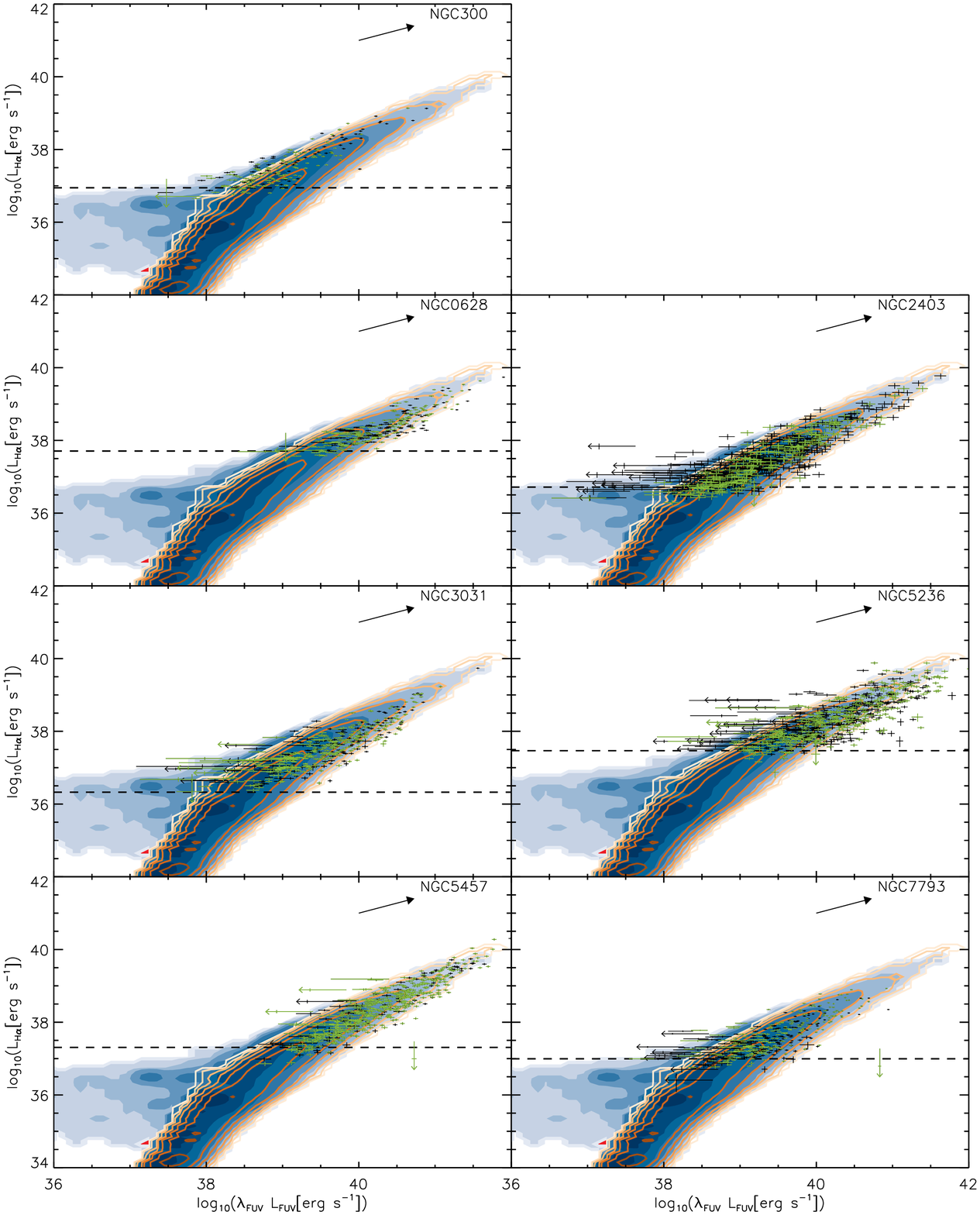}
\end{minipage}
\caption[Results: rest of the sample]{As in Figure \ref{7gals_HvF}., but with \citet{Calzetti2007} attenuation recipe applied in the place of \citet{Hao2011}. Note the presence of larger scatter at high luminosities.}
\label{7gals_HvF_Calzetti}
\end{figure*}

\begin{figure*}
\begin{minipage}{180mm}
\centering
\includegraphics[width = 1.0\columnwidth]{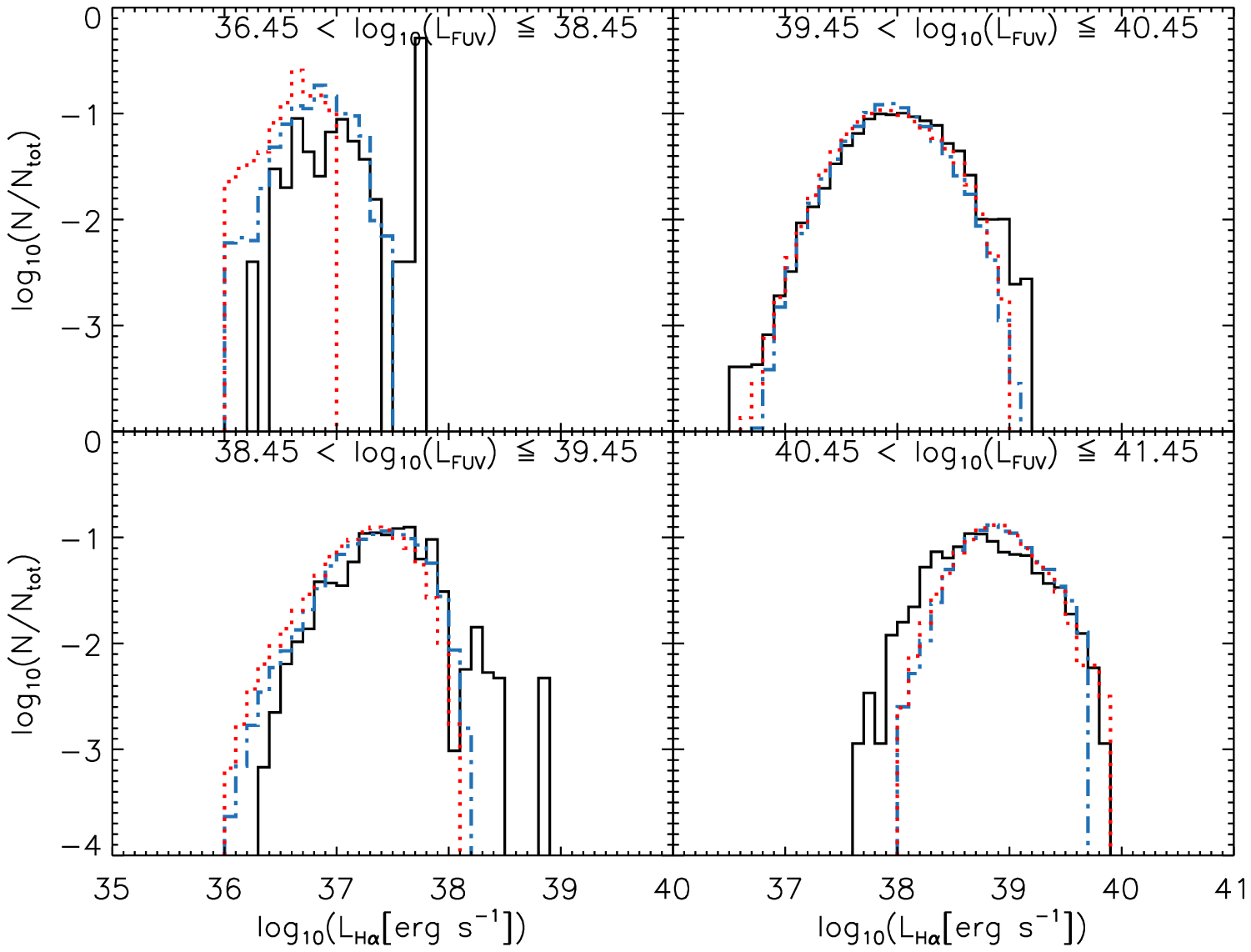}
\end{minipage}
\caption[\Ha luminosity stack]{As in Figure \ref{Ha_hist_stack}., but with \citet{Calzetti2007} attenuation recipe applied in the place of \citet{Hao2011}.}
\label{Ha_hist_stack_calzetti}

\end{figure*}

\begin{figure*}
\begin{minipage}{180mm}
\centering
\includegraphics[width = 1.0\columnwidth]{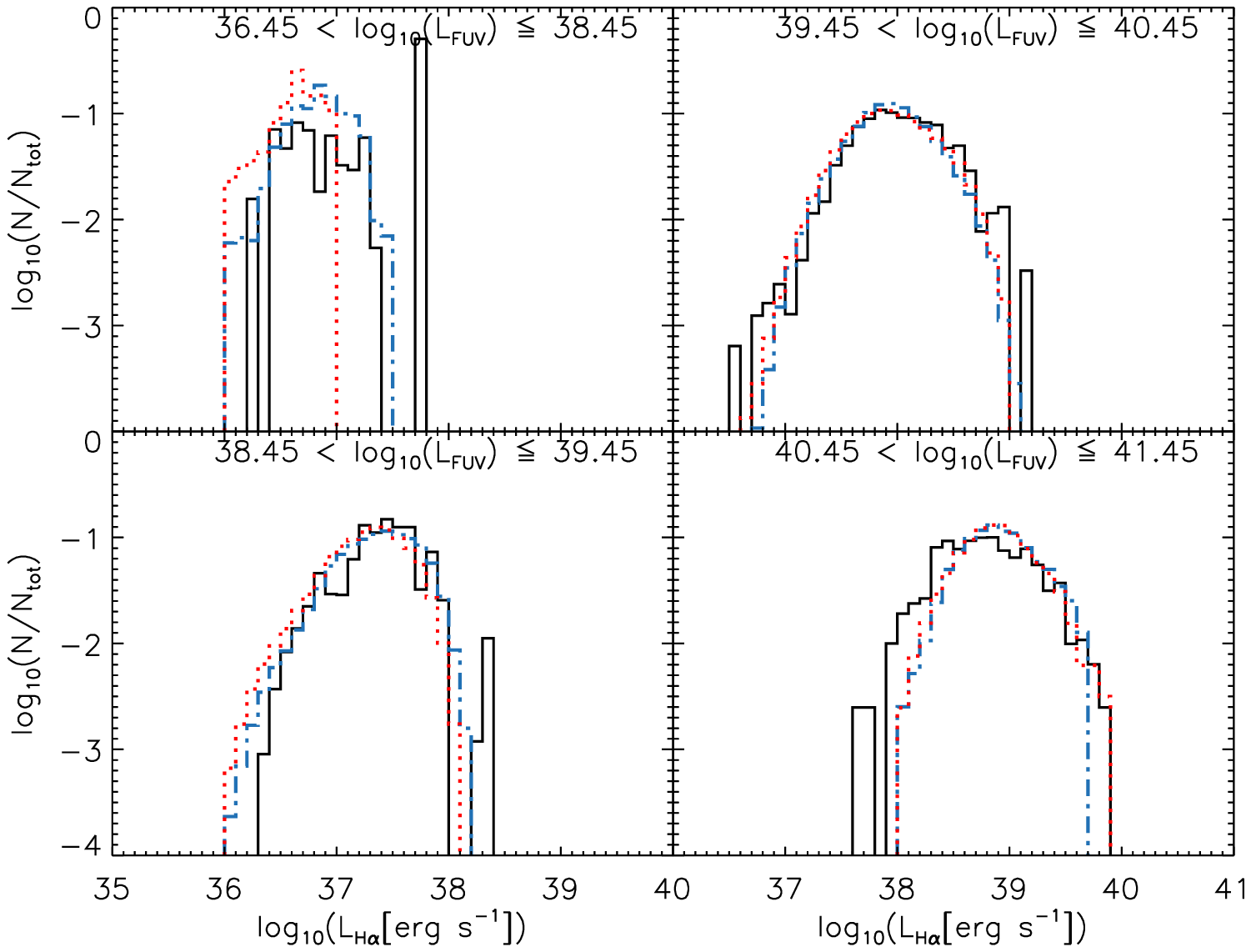}
\end{minipage}
\caption[\Ha luminosity stack - handpicked]{As in Figure \ref{Ha_hist_stack_reghand}., but with \citet{Calzetti2007} attenuation recipe applied in the place of \citet{Hao2011}.}
\label{Ha_hist_stack_reghand_calzetti}

\end{figure*}

\end{document}